\documentclass{llncs}

\usepackage[round]{natbib} 
\usepackage{graphicx} 
\usepackage{amsmath,amssymb,amsfonts}
\usepackage{float}
\usepackage{enumerate}
\usepackage[margin=1 in]{geometry}
\usepackage{comment}
\usepackage{xcolor}
\usepackage{pgfplots}
\usepackage{subfig}
\usepackage{algorithm}
\usepackage{algorithmic}

\usepackage[colorlinks=true, linkcolor=blue, citecolor=blue]{hyperref}

\usepackage{multirow, tabularx, booktabs}
\newcommand\setrow[1]{\gdef\rowmac{#1}#1\ignorespaces}
\newcommand\clearrow{\global\let\rowmac\relax}
\clearrow
\begin{document}

\title{A novel robust mixed integer linear programming model for index tracking problem under no rebalancing: heuristic optimization approach}

\author{
Danial Ramezani \inst{1} \and
Mostafa Abouei Ardakan \inst{2}  \and
Mohamadreza Dehghani  Ahmadabad  \inst{3} 
}
\institute{
MSc. Student, Department of Industrial Engineering, Faculty of Engineering, Kharazmi University, Tehran, Iran.\\
\email{danialramezani@khu.ac.ir}
\and
Associate Prof., Department of Industrial Engineering, Faculty of Engineering, Kharazmi University, Tehran, Iran\\
\email{abouei@khu.ac.ir}
\and
Assistant Prof., Department of Financial Management, Faculty of Financial Sciences, Kharazmi University, Tehran, Iran\\
\email{mr.dehghani@khu.ac.ir}
}

\maketitle              
\begin{abstract}
Passive management has increasingly won popularity over the past few years because of its 
advantages, such as lower management fees and transaction costs. Index tracking 
endeavors to reproduce the performance of an index with smaller sets of assets. In this paper, a novel 
formulation is proposed that is not only more robust than the existing ones but also performs better 
on out-of-sample data and tracks indices over long periods without any considerable
deviation or the need for rebalancing. Solving index tracking problems in a polynomial time is a 
challenging task due to their NP-hard nature. To address this issue, a novel heuristic based on 
metaheuristic algorithms and local branching is also developed to solve the proposed model. The
heuristic enjoys not only the exploration capabilities of a genetic algorithm but the characteristics 
of local search algorithms as well. The data from the OR library is used to verify the capabilities 
of the proposed heuristic in comparison with commercial solvers. Results indicate that not only is 
the heuristic able to converge to optimal solutions for not-so-large problem sizes, but the portfolios 
it generates also outperform those yielded by commercial solvers in terms of both in-sample and 
out-of-sample data. 

\end{abstract}

\section{Introduction}
Fund management, alternatively called ‘asset management,’ involves making decisions on 
behalf of clients to manage their capital toward profits based on their desired risk level. The 
management styles commonly employed in the relevant financial institutions may be clustered 
under the two general groups of active and passive management.

Active fund managers proactively make decisions on changing the composition of the portfolio 
and buying or selling assets to gain profits. Indeed, it is the objective of these funds to outperform 
the market. To achieve this goal, the managers draw upon their expertise, knowledge, and experience 
in selection of securities, overview of the economy, and market timing, among others. Compared 
to passive funds, this management style, however, suffers from such drawbacks as higher 
management fees, regardless of the profitability of the manager, and higher portfolio turnovers and 
transaction costs. Obviously, the cumulative costs may, at times, ruin the marginal profits made.

Passive fund management, also known as passive investment or indexing, is based on the 
belief that it is not possible to outperform the market in the long run. This belief is crystalized in
the celebrated saying that ‘time in the market is more important than market timing’. Indeed, it is 
the goal of this investment style to create portfolios of constituent assets that replicate the return 
of the benchmark (market index). In this situation, a simple solution that comes to mind is full 
replication. However, buying or selling some of the companies, especially small ones, in the 
market might run the risk of liquidity since the costs associated with transactions, monitoring, and 
rebalancing can be huge enough to make this strategy impractical in real-life applications. The
strategy often adopted by managers to reduce the costs and the liquidity risks is called ‘index 
tracking’ (IT) whereby a subset of assets in the market is selected to make a portfolio capable of
tracking the return of the market. Due to its wide use by fund managers, IT has attracted a lot of 
interest among researchers in recent decades. Another approach entertained by fund managers is
enhanced IT or enhanced indexation defined as outperforming a given index (benchmark) while 
not deviating much from it. In this approach, managers look for portfolios that provide excess 
returns within their tolerable risk profile

From among the different modeling frameworks available for addressing IT problems, the 
mathematical modeling approach is the most common in the literature (\cite{silva2023systematic}). In this framework, the problem is defined as a combinatorial optimization one resulting from the selection of a subset of 
assets. Such problems are NP-hard in nature to which a vast amount of research efforts has been
dedicated in the quest of heuristics for their solution. The logic underlying the optimization approach 
is the assumption that ’a tracking portfolio is a good solution to the problem in out-of-sample data 
if it had good tracking records in the past and performed well on in-sample data’. Put differently, 
it is generally assumed that a portfolio with a low tracking error in the past will continue to have a 
low tracking error in the future.

\subsection{Literature review}

The literature on tracking error measurement is extensive and diverse in scope.

\cite{roll1992mean} studied 
the efficient frontier of the IT problem by minimizing the tracking error variance subjected to some 
level of expected return. He showed that constraining beta in some cases could lead to better 
solutions. \cite{jansen2002optimal} considered the number of assets chosen and incorporated it into the optimization model as an objective function. \cite{jorion2003portfolio} argued that it does not suffice to consider solely the tracking error; rather, the author suggested that more efficient solutions could 
be obtained if the variance of the tracking portfolios were constrained to values below that of the 
benchmark.

\cite{beasley2003evolutionary} proposed a mathematical model for index tracking and enhanced indexation in which transaction costs and a cardinality constraint are considered. The authors subsequently 
developed a heuristic-based on population for its solution, tested their approach using five major 
world market indices, and recorded the computation times for the various parameters in their 
model. \cite{derigs2004local} designed a 2-phase simulated annealing heuristic in which the return 
of the assets was assumed to have a linear combination of macroeconomic factors. They validated 
their heuristic by applying it to a real-world case study. \cite{gaivoronski2005optimal} studied the effects
of various risk and target measures introduced into IT mathematical models as well as those of 
their threshold policy on rebalancing portfolios. They tested their strategies on stock data from 
the Oslo stock exchange. \cite{alexander2005indexing} compared tracking error variance-minimizing models with cointegration-based ones for tracking indices to find that both model types yielded
similar results. \cite{oh2005using} employed a genetic algorithm (GA) for the Korea Stock Price Index 
under various market regimes and studied the tracking error of portfolios under different scenarios. \cite{maringer2007index} developed a differential evolution algorithm and tested it on the Dow 
Jones Industrial Average Index. They suggested that it would be inappropriate to consider 
information from too far in the past for in-sample optimization since outdated observations could 
be irrelevant now. These authors claimed further that larger portfolios would have better tracking 
records and that a portfolio with half the available assets would be large enough to yield
satisfactory tracking results. 

\cite{canakgoz2009mixed} presented a three-stage solution with the first two stages dedicated 
to the estimation of the desirable regression’s beta and alpha slopes of the portfolio while the last 
phase would deal with minimizing transaction costs subjected to the slopes already estimated. 
Another facet of the study involved the study of the relationship between the number of assets in a
portfolio and the out-of-sample alpha and beta factors. Although the results revealed no clear 
relationship between the number of assets and alpha, the out-of-sample’s beta factor and number of 
assets showed a strong correlation up to a given threshold. \cite{ruiz2009hybrid} presented a hybrid GA for the IT problem using quadratic programming as a fitness function 
to determine the weights of the assets. \cite{krink2009differential} proposed a hybrid differential evolution 
algorithm that utilized a combinatorial search operator. They tested their results on an artificial
index constructed based on market values of the assets in the nikkei market at the beginning of 
the period. 

\cite{stoyan2010two} formulated an index tracking stochastic mixed integer optimization 
programming problem and developed a novel way for its decomposition. Instead of formulating 
IT to have similar returns relative to the index,  \cite{guastaroba2012kernel} proposed a mixed-integer linear programming model that tracks market index value instead of the returns; the authors then designed a novel heuristic to solve the problem and ran the heuristic on instances of OR-library (\cite{beasley1990or}) data. \cite{chen2012robust} proposed a robust formulation for IT and tested their model using the S\&P 100 index. \cite{wang2012mixed} added a CVaR constraint to the problem formulation and 
tested their model on OR-library indices. Numerical analysis showed that even though adding 
CVaR to the problem had no impact on tracking error for positive returns, it was able to limit the 
downward returns of the tracking portfolio. \cite{ni2013stock} considered tracking error, excess 
return, and tracking error variance as objectives and proposed a heuristic to solve the problem. \cite{scozzari2013exact} proposed a mixed integer quadratic programming and incorporated into it the 
UCITS rules (i.e., the rules required for fund management in the European Union). \cite{bruni2015linear} proposed a formulation for enhanced indexation in which excess return is maximized while
maximum underperformance is minimized. They tested their formulation on a publicly available 
benchmark. \cite{guastaroba2016linear} used the omega ratio as the objective to formulate the enhanced IT 
problem and incorporated real-world constraints such as cardinality and buy-in threshold. \cite{filippi2016heuristic} considered excess return and tracking error as objectives and adopted an approach that 
would combine the kernel search heuristic using the $\epsilon$-constraint method.

\cite{strub2018optimal} suggested that IT formulations that track returns of an index focus on 
shape and may have large distances in a specific period. On the other hand, formulations that track 
index value are designed to capture a close distance between the portfolio and the index but not
the shape. They, therefore, proposed a new formulation that accounted for these shortcomings in 
three financing settings for transaction costs.

\cite{sant2017index} employed GA to develop a hybrid heuristic and provided the related 
computation results for S\&P 100, DAX, and FTSE. \cite{wu2017constrained} proposed mathematical models with 
turnover, trading, and sector diversification constraints. They additionally developed heuristics to 
create feasible initial solutions and adopted Lagrangian and semi-Lagrangian methods to obtain 
the final solution. Strub and Trautmann \cite{strub2019two} implemented a two-stage method whereby a solution 
is found by GA in the first stage, and the best solution is found via a local search heuristic in the 
second stage. They also proposed a new formulation for UCITS rules. \cite{sehgal2019enhanced} proposed a model for enhanced indexation based on weighted conditional value at risk and 
compared the results with those obtained from three different models in various market conditions 
to create a fair situation. \cite{kaucic2020polynomial} considered downside deviation from the benchmark, upside 
potential profit, and modified sharp ratio as their objectives; combined them using goal 
programming; and solved the problem on Euro Stoxx 50 using an improved version of the particle 
swarm optimization along with a novel constraint handling technique.

\cite{gnagi2020tracking} studied enhanced IT models and suggested that considering tracking error 
variance instead of mean absolute deviation could lead to better out-of-sample performance. They 
proposed new formulations with various real-world constraints for both objectives and designed a 
new meta-heuristic for acquiring solutions. \cite{sant2020solving} proposed a mixed-integer non-linear programming formulation for the IT problem by using cointegration methodology and 
solved it using a branch-and-cut algorithm. Their results indicated that their model provided
portfolios with lower turnover and transaction costs. \cite{sant2022risk} studied the effects of 
adding seven risk measures as constraints to the IT optimization problem. They also developed 
metrics to determine the quality of each risk measure.

\cite{torri2023penalized} utilized penalization to control the deviation of portfolio weights from the 
benchmark. In contrast to the conventional strategies, their methodology placed a greater emphasis 
on assessing the portfolio’s risk rather than on its risk relative to the market and reduced turnover. \cite{silva2023using} developed two new evolutionary algorithms and compared the 
performance of models that use historical data and those that use data generated by generative 
adversarial networks.

\cite{vieira2023liquidity} conducted a comprehensive study of liquidity in IT problems and proposed two 
distinct formulations for liquidity in the IT model. The first approach that is based on weighted sum of liquidities revealed that
number of assets in the portfolio decreases with increasing restrictions on liquidity while their second approach does not have this feature. \cite{anis2023risk} proposed a novel formulation that evenly divided
the tracking portfolio risk among all the sectors. They also proposed heuristics to solve the 
problem. \cite{beraldi2022enhanced} studied the enhanced IT problem using chance constraints and 
presented a framework in which the tracking portfolio’s return was enforced to surpass the 
benchmark with a high probability. \cite{silva2024enhanced} adopted the greedy randomized adaptive search 
procedure (GRASP) in IT and showed that their proposed GRASP was able to provide almost as good-quality solutions as did CPLEX but in a shorter time. \cite{chang2025index} showed that rigidity among index-tracking investors around MSCI index reconstitutions creates predictable price pressure and trading imbalances that generate exploitable arbitrage opportunities, particularly between announcement dates and the trading day preceding the effective date. \cite{zhang2025index} proposed a sparse Bayesian regression framework combined with collaborative neurodynamic optimization to construct sparse and enhanced index-tracking portfolios that outperform traditional methods in predictability, consistency, sparsity, and profitability.\cite{grassetti2025optimizing}  developed an index-tracking framework that integrates Random Matrix Theory with network-based eigenvalue centrality to build compact, representative portfolios that achieve low tracking error and robust performance across varying market regimes with high computational efficiency. \cite{xu2026network} proposed network-based index tracking models—ITN and its adaptive version ITNA—that incorporate asset dependency structures through community and centrality constraints, demonstrating improved adjusted returns and lower tracking errors compared to traditional index-tracking approaches. \cite{talaei2025robust} developed a clustering-based index tracking model for the Tehran Stock Exchange that incorporates turnover and buy-in constraints and applies possibilistic chance-constrained and robust possibilistic programming to handle parameter uncertainty, demonstrating improved robustness and practical performance under varying market conditions. \cite{zapata2025enhancing} showed that deep learning models—specifically autoencoders and variational autoencoders—can construct sparse, cardinality-constrained portfolios that closely replicate the performance of the Nasdaq-100 with reduced tracking error and competitive cumulative returns. \cite{fieberg2025enhancing} introduced a cardinality-constrained mixed-integer optimization framework that integrates characteristic-based factor models to reduce estimation uncertainty, demonstrating consistently lower tracking errors and more robust index-tracking performance than traditional linear and quadratic approaches across markets and transaction cost settings. \cite{cesarone2025benchmark} proposed a benchmark-asset principal component factorization method that uses spectral decomposition of benchmark–asset covariance matrices to construct small, low-turnover tracking portfolios with competitive out-of-sample performance and substantially greater computational efficiency than traditional optimization-based approaches. \cite{dhingra2026comprehensive} provided a comprehensive review of optimization-, statistical-, and machine learning–based approaches to financial index tracking and, using empirical evidence from the S\&P 500, show that tracking error volatility models offer the most precise replication, convex cointegration achieves the best risk–return trade-off, and deep neural networks deliver competitive performance with low turnover and high efficiency. \cite{wang2026exact} developed an exact algorithm for cardinality-constrained sparse index tracking that incorporates investor preferences through fuzzy preference relations and additive consistency, combining matrix decomposition, supergradient methods, and branch-and-bound to improve computational efficiency and deliver superior risk–return and tracking performance compared to existing mixed-integer approaches. \cite{sammon2026index} showed that value-weighted index rebalancing in response to issuance, buybacks, and IPOs effectively induces systematic market timing that generates negative factor exposures and a 46–69 basis point annual performance drag, and they demonstrate that less frequent or delayed rebalancing rules can closely track the market while materially improving returns and reducing trading costs.

\subsection{Contribution of the current paper}
The contribution of this paper is twofold. First, a novel mixed-integer linear programming 
formulation is proposed that is much more robust than the traditional IT models in that it not only 
produces better out-of-sample results but also tracks indices with more precision over long periods
of time without the need for rebalancing. Second, given the fact that mixed-integer programming 
models, especially IT ones are NP-hard (\cite{de2024assessing}), a powerful heuristic is herein proposed that, compared to 
commercial solvers, is capable of providing solutions of equal or even superior quality in a shorter 
time. We merged the exploration capabilities of the genetic algorithm and the local search one of 
local branching methods to develop a unique genetic algorithm with local branching (or GALB for 
short). The algorithm is found superbly capable of converging to global optima, thereby
outperforming commercial solvers. 
\subsection{Structure of the paper}
The rest of this paper is organized as follows. Section \ref{Mathematical model} provides information regarding 
existing formulations and presents the novel robust formulation proposed. Section \ref{Solution approach} presents
GALB. The experimental results obtained from the comparisons of the proposed model and the 
conventional ones with respect to model power and robustness as well as those indicating the 
superiority of our heuristic approach over exact solvers are presented in Section \ref{Experimental results}. Finally,
concluding remarks are provided in Section \ref{Conclusion}.

\section{Mathematical model}
\label{Mathematical model}
This section reviews the currently existing IT formulations.
The first model to consider is that described in \cite{cornuejols2006optimization} that is expressed by
Model 1 below.\\

\textbf{Model 1:}

Assume a market with $N$ assets with $\rho_{ij}$ expressing the similarity between assets $i$ and $j$. A widely accepted metric in the literature for expressing similarity between assets is their correlation.

\begin{flalign}\label{Eq:Obj1}
Z = \max\  \sum_{i=1}^{N}\sum_{j=1}^{N} \rho_{ij}x_{ij} && 
\end{flalign}

Subject to:

\begin{flalign}\label{Eq:cardinality1}
\sum_{j=1}^N y_j = q &&
\end{flalign}

\begin{flalign}\label{Eq:represent1}
\sum_{j=1}^N x_{ij} = 1&&  \forall i=1,...,N
\end{flalign}
\begin{flalign}\label{Eq:xBeOneOrZero1}
x_{ij} \leq y_j&&  \forall i,j=1,...,N
\end{flalign}
\begin{flalign}\label{Eq:Domain1}
x_{ij}, y_j = \{0,1\}&&  \forall i,j=1,...,N
\end{flalign}

Where $y_j$ is a binary variable that is equal to unity if asset $j$ is in the tracking portfolio and zero,
otherwise. Another binary variable is $x_{ij}$ that is equal to 1 if asset $j$ is most similar to asset $i$ and zero, otherwise. It basically shows whether asset $j$ can be a representative of asset $i$ in the portfolio.

Equation (\ref{Eq:Obj1}) above is the objective function that maximizes the similarity between $N$ stocks 
and their representatives in the fund. Equation  (\ref{Eq:cardinality1}) forces the tracking portfolio to have $q$ assets, while Equation (\ref{Eq:represent1}) ensures that each asset $i$ has only one representative in the selected portfolio. Equation (\ref{Eq:xBeOneOrZero1}) guarantees that asset $i$ can be represented by $j$ if $j$ is chosen to be in the tracking portfolio. Finally, the last equation defines the domains of the variables.

This model only selects the best $q$ assets but does not directly assign weights to them. The 
weight chosen by the model for asset $j$ is defined as follows:

\begin{flalign}\label{Eq:weight1}
w_j = \sum_{i=1}^N V_i x_{ij} &&
\end{flalign}

where $V_i$ is market capitalization of asset $i$. The appropriate capital to invest in asset $j$ can be 
determined by simply normalizing $w_j$s.

This approach has its own drawbacks. A quite obvious one is the use of correlation data of assets 
as a similarity measure since correlation between assets can be noisy and constantly changing 
across the changing market conditions. This issue is further compounded by the fact that there exist situations in reality where people have limited data or the data are invalid due to unexpected events in society (\cite{yang2023new}). On the one hand, as also observed by \cite{maringer2007index}, we cannot look too far back in the past to collect data for in-sample optimization and
on the other hand, failing to capture correlation data in various market conditions might result in poor 
out-of-sample results. To address this drawback, \cite{chen2012robust} proposed a robust version of Model 1 by adopting the optimization approach due to \cite{bertsimas2003robust} and converted the equation (\ref{Eq:Obj1}) to the following equation:

\begin{flalign}\label{Eq:convertObj1}
Z = \max_{x \in X} \sum_{i=1}^N \sum_{j=1}^N \{\rho_{ij}x_{ij} + \min_{\{S|S \subseteq N,|S| \leq \Gamma \}} \{\sum_{i,j \in S} -d_{ij}x_{ij} \} \} &&
\end{flalign}

Where $\rho_{ij}$ is the nominal or value expected value for the correlation between $i$,$j$ and $d_{ij}$ being a non-negative downside deviation for it, $\Gamma$ represents the number of coefficients that 
take their worst possible value, and $S$ is some subset of $N$ such that the highest $\Gamma$ deviations are realized. Thus, Model 1 can be transformed into its robust form as Model 2:

\textbf{Model 2:}

\begin{flalign}\label{Eq:obj2}
Z = \max \sum_{i=1}^N \sum_{j=1}^N \{\rho_{ij} x_{ij} - v_{ij} \} -\Gamma \theta &&
\end{flalign}

Subject to:

(\ref{Eq:cardinality1})(\ref{Eq:represent1})(\ref{Eq:xBeOneOrZero1})(\ref{Eq:Domain1})

\begin{flalign}\label{Eq:ro2}
v_{ij} + \theta \geq d_{ij}x_{ij} && \forall i,j =1,...,N
\end{flalign}

\begin{flalign}\label{Eq:roDomain2}
\theta , v_{ij} \geq 0 && \forall i,j =1,...,N
\end{flalign}

However, ignoring the proportional weights of assets in the model overlooks real-world 
constraints and might lead to inaccurate representations.
Model 3 (\cite{silva2023systematic2}) is yet another IT model that is well-established in the literature.\\

\textbf{Model 3:}

Assume we have T periods and want to find a portfolio with the least deviation from the return 
of the benchmark while not violating some constraint related to holding different assets. This 
model has gained popularity among researchers since it considers the weights of the assets and is, 
therefore, capable of considering more complicated real-world constraints and risk measures.

\begin{flalign}\label{Eq:obj3}
Z = \min \dfrac{1}{T} \sum_{t=1}^T |I_t - \sum_{i=1}^N {r_{it}w_i}| &&
\end{flalign}
\begin{flalign}\label{Eq:sum3}
\sum_{i=1}^N w_i = 1 &&
\end{flalign}
\begin{flalign}\label{Eq:cardinality3}
\sum_{i=1}^N z_i = K &&
\end{flalign}

\begin{flalign}\label{Eq:boundary3}
\epsilon_i z_i \leq w_i \leq \delta_i z_i && \forall i = 1,...,N
\end{flalign}
\begin{flalign}\label{Eq:domain3}
w_i \geq 0 , z_i = \{0,1\} && \forall i = 1,...,N
\end{flalign}

Where $w_i$ is the weight of asset $i$, $z_i$ is a binary variable that is 1 if asset $i$ is chosen in the tracking portfolio and 0 otherwise. $I_t$ is return of the benchmark in time $t$, $r_{it}$ is return of the asset $i$ at time $t$. Equation (\ref{Eq:obj3}) minimizes the mean absolute deviation of the portfolio's return and index return. Equation (\ref{Eq:sum3}) ensures that all the capital is invested. Cardinality constraint is defined as (\ref{Eq:cardinality3}). Minimum and maximum amount of holding of different assets are determined by (\ref{Eq:boundary3}) such that if asset $i$ is selected, its weight in the portfolio is not less that $\epsilon_i$ and does not exceed $\delta_i$. Finally, equation (\ref{Eq:domain3}) prohibits short selling and defines the domain of the variables.

It is fair to say that although mean absolute deviation is popular, there exist other indicators 
such as mean squared error and root mean squared error that are also popular and widely used in 
the literature. However, the advantage of mean absolute error, compared to its counterparts, is its 
ease of linearization. The linear form of Model 3 may be rewritten as follows:

\begin{flalign}\label{Eq:objLinear3}
Z = \min \dfrac{1}{T} \sum_{t=1}^T( u_t + d_t) &&
\end{flalign}
Subject to:

(\ref{Eq:sum3})(\ref{Eq:cardinality3})(\ref{Eq:boundary3})(\ref{Eq:domain3})

\begin{flalign}\label{Eq:objDU3}
u_t - d_t = I_t - \sum_{i=1}^N {r_{it}w_i} &&
\end{flalign}
\begin{flalign}\label{Eq:domainDU3}
u_t , d_t \geq 0 && \forall t=1,...,T
\end{flalign}

\subsection{Our formulation:}

In this Subsection, the IT formulation proposed in this study is introduced as Model 4.

\textbf{Model 4:}

\begin{flalign}\label{Eq:obj4}
Z = \min \dfrac{\xi}{T}\sum_{t=1}^T(u_t + d_t) -  \frac{\sum_{i=1}^N \sum_{j=1}^N \{\rho_{ij} x_{ij} - v_{ij} \} -\Gamma \theta }{\psi} && 
\end{flalign}

Subject to:

(\ref{Eq:represent1})(\ref{Eq:Domain1})(\ref{Eq:ro2})(\ref{Eq:roDomain2})(\ref{Eq:sum3})(\ref{Eq:cardinality3})(\ref{Eq:boundary3})(\ref{Eq:domain3})(\ref{Eq:objDU3})(\ref{Eq:domainDU3})

\begin{flalign}\label{Eq:onlyOneRep4}
x_{ij} \leq z_j && \forall i,j=1,...,N
\end{flalign}

Where $\psi$ is the objective value of Model 1 and $\xi$ is a positive coefficient. $\psi$ and $\xi$ are used to scale both terms in the objective between 0 and 1.

Unlike the traditional models in which market capitalization is used for weighting assets, the 
present one is capable of selecting a portfolio with $K$ assets that
can represent market in the worst possible condition and providing optimal weights for it. As another advantage of the model real-world 
constraints such as cardinality and boundary constraints can be added to it.

Relying on in-sample data to optimize a model toward good performance on out-of-sample data 
could be tricky in our case since correlation between assets is used in the objective function and it 
is known that correlation among assets might have quite different values under different market 
regimes. The next step is, therefore, to determine the downside deviation of correlations among
assets. To accomplish this, different periods with fixed lengths were considered in a market and 
the correlations of the assets were recorded for those periods. Finally, the standard deviations of 
the correlation values were calculated and recorded as downside deviation.
\section{Solution approach:}
\label{Solution approach}
This Section describes the approach adopted to solve the proposed model. As already 
mentioned, IT models are NP-hard such that acquiring optimal solutions in an acceptable 
timeframe becomes impossible as problem size increases. To address this issue, we developed a 
novel heuristic based on evolutionary algorithms and local branching (GALB), a detailed 
description of which is provided in the Subsection (\ref{Local branching}). Genetic algorithm is used as an evolutionary 
algorithm and functions as the primary search algorithm. If some specified conditions are met after 
several search iterations, the local branching is performed on the best solution using some special 
sets and settings. GALB is not only capable of converging to the optimal solution when the problem 
size is not too large but is also able to produce better results in lower time limits than do exact 
solvers in mixed integer programming models. In the following subsections, various components 
of GALB are briefly elucidated.

\subsection{Genetic algorithm:}
Genetic algorithm (GA) is an optimization technique that belongs to the class of evolutionary 
algorithms and is greedy in nature inspired by the concept of natural selection. The main idea 
underlying GA is to generate iteratively a population of candidate solutions of multiple 
generations, preserving high-quality solutions and eliminating the poor ones by implementing such 
operators as crossover and mutation. In this way, the fittest individuals (i.e., those that have higher 
values of the objective function) are preserved to mate with each other in order to produce better 
generations. Since GA is an approximate optimization technique, it is likely that it cannot converge 
to global optima. Algorithm \ref{GA} demonstrates a simple GA:

\begin{algorithm}[H]

\caption{Genetic Algorithm}
\begin{algorithmic}
\label{GA} 
\STATE Initialize a random population of solutions of size $p$;
\WHILE{Stopping criteria is not true}

\STATE Create offspring by applying crossover to the population of candidate solutions;
\STATE Create offspring by applying mutation to the population of candidate solutions;
\STATE Add generated offspring to the pool of current solutions
\STATE Determine the fitness of the generated population;
\STATE Evaluate fitness of all solutions;
\STATE Preserve the best $p$ individuals and eliminate rest of the solutions;
\STATE Save the best solution and its fitness value

\ENDWHILE
\STATE Return the best solution and its fitness value

\end{algorithmic}
\end{algorithm}

\subsection{Calculating fitness function:}

There exist several ways to determine the fitness of each solution in GA. One straightforward 
way is to use an exact solver as a fitness function in GA. Nevertheless, this method has the obvious 
drawback of an immense additional time complexity added to the algorithm. Another possible 
option is to calculate the objective function directly and define penalty functions for constraint
violation. The disadvantage of this technique is that probably more time or iterations will be needed
to find a good solution or even a feasible one, thereby harming the good performance of the 
algorithm. In addition, as is the case with exact solvers, the fitness function design should be based 
on the consideration that some of the solutions have some degree of relative 
superiority over the others. For illustration, let us suppose that there exist portfolios A and B that 
are presented as solutions by an exact solver. If A is better than B in terms of the objective function 
value, the fitness function designed should also be able to reflect this superiority. However, it is 
clear that the chances are high that the reflection of this supriority will face problems if this second 
technique is implemented.

Assuming a market with $N$ assets, our encoding system is composed of vectors of size $N$ where each element of the vectors determines the weight of each asset. Algorithm \ref{fitness} below captures
our approach to designing a fitness function that neither is as time-consuming as using an exact 
solver, nor does it cause the superiority problem.

\begin{algorithm}[H]

\caption{Fitness function}
\begin{algorithmic} 
\label{fitness}
\STATE Let $P$ be the population of solutions; 
\STATE Let $W_{ij}$ be $\rho_{ij} - d_{ij}$ 
\STATE Let $N$ be the universe of assets;
\STATE $D_i$ = $\{d_i \hspace{2mm}|\hspace{2mm} d_i = \arg\max P_i,|d_i| = K \}  \forall i \in P$; // Decoding phase
\STATE $E = \{N\} - D$;
\FOR {$i=1$ to $D$}
\STATE $R = 0$;
\STATE temp = $\{\}$;
\FOR {$e \in E $}
\STATE $M_x  = \max(W_{eD_i}) $; // maximum value of $e$ and elements in $D_i$
\STATE $R = R + M_x$;
\IF {$\Gamma = N$}
\STATE Continue; 
\ELSE
\STATE $y = \{y\hspace{2mm}|\hspace{2mm} W_{ey} = M_x \}$;
\STATE Add $d_{ey}$ to temp; 
\ENDIF
\ENDFOR
\IF{$\Gamma \neq N$}
\STATE tempS = Sorted temp set in an ascending order;
\STATE $\displaystyle R = R + \sum_{i=1}^{|temp|+K-\Gamma}(\text{tempS}_i)$;
\ENDIF
\STATE $F_i = \dfrac{\xi}{T} \sum_{t=1}^T |I_t - \sum_{i=1}^N {r_{it}w_i}| - \dfrac{R}{\psi}$;
\ENDFOR
\end{algorithmic}
\end{algorithm}

Initially, chromosomes are decoded into numbers of the assets. To calculate the second term in the equation (\ref{Eq:obj4}) described above, all the worst possible values are assigned to the second term in this 
equation. The maximum value of $W_{ij}$ of an asset that is not in the portfolio and all the assets in 
the portfolio are recorded. In the next step, if some parameters are allowed not to take the worst 
possible value, the downside deviation of values with the lowest $d_{ij}$ is added to make them 
nominal values. Using the local branching and repairing mechanism as well as the fitness function 
thus obtained, a quite decent tool is now available to compare solutions.

\subsection{Local branching}
\label{Local branching}
Put simply, the rationale behind local branching is to search the space of best solutions found 
in the quest of the better ones in the neighbourhood. It is based on hamming distance of solutions; the hamming 
distance between two vectors of equal length is the number of positions at which the corresponding 
symbols are different. For example, given the two solutions A = [1,0,0,1] and B = [1,1,0,0], the 
distance between A and B determines the number of flips required to derive solution B from 
solution A. In this case, the first and third bits are identical and no change is, therefore, needed. To 
create B from A, the second bit has to change to 1, and the last one has to change to 0; since two 
changes are required, the distance between these two will be two. The very same logic is adopted 
in local search as mathematically represented in Model 5.

\textbf{Model 5:}

\begin{flalign}\label{Eq:obj5}
Z = \min \dfrac{\xi}{T}\sum_{t=1}^T(u_t + d_t) -  \frac{\sum_{i=1}^N \sum_{j=1}^N \{\rho_{ij} x_{ij} - v_{ij} \} -\Gamma \theta }{\psi} && 
\end{flalign}

Subject to:

(\ref{Eq:represent1})(\ref{Eq:Domain1})(\ref{Eq:ro2})(\ref{Eq:roDomain2})(\ref{Eq:sum3})(\ref{Eq:cardinality3})(\ref{Eq:boundary3})(\ref{Eq:domain3})(\ref{Eq:objDU3})(\ref{Eq:domainDU3})(\ref{Eq:onlyOneRep4})

\begin{flalign}\label{Eq:LB1}
a \leq \sum_{j \in J}(1-z_j)+\sum_{i \in I \setminus J} z_i \leq b && 
\end{flalign}
\begin{flalign}\label{Eq:LB2}
\sum_{i \in N \setminus I} z_i = 0 &&
\end{flalign}

Model 5 is equivalent to Model 4 except that it contains two additional constraints used for local 
branching. Equation (\ref{Eq:LB1}) determines the number of flips in the current best solution $J$ and the considered assets $I$. Equation (\ref{Eq:LB2}) ensures that assets other than set $I$ are not selected. To create a set of assets in the Local branching algorithm ($S_s$), relaxation of Model 4 is solved. In 
the next step, a percentage of the universe of assets in the market (designated by $L$) is determined 
to select the subsequent values. The next top $L$ assets with the highest $z$ and $w$ are picked. Consider $W$, which is the matrix of the worst possible correlation values between the assets. The $L$ largest values of the mean of each row of that matrix is calculated and added to $S_s$.  Thus, a 
promising set is created that will be used in the local branching process. One key note to mention
is that even values should be selected for $a$ and $b$  since the number of assets in the portfolio 
should be fixed for the present case; odd values will, therefore, produce infeasible solutions.
Another point of importance is that the minimum number of flips that keep the portfolio feasible 
is two, while the maximum number of bit flips allowed without producing a totally new portfolio 
(that is, one that contains at least one asset from the original portfolio) is  $2(K-1)$). In this 
situation, if the portfolios are of size 3 ($K=3$), then the maximum distance that can be searched 
is achieved by changing four bits (i.e., setting two assets from the original portfolio to zero and 
two new ones to one); this is because five flips would produce infeasible solutions and six flips 
would require all ones in the portfolio to be set to zero. For our present purposes, we opt for smaller 
and random but more promising sets to consider and search larger distances, rather than having 
larger sets but searching within smaller distances. Algorithm \ref{Local branching-alg} captures the local branching 
heuristic used in this paper.

\begin{algorithm}[H]

\caption{Local branching}
\begin{algorithmic} 
\label{Local branching-alg}
\STATE Let $rep$ be the number of time that the solution is not improved by the local branching process;

\STATE Let $c_1,c_2 \in \mathbb{R}, (0 \leq c_1,c_2 \leq 1),c_2 \geq c_1, c_1 \leq 0.5,c_2 \geq 0.5 $;
\STATE Let $\alpha,\beta \in \mathbb{Z}^+ , \beta \geq \alpha$ ;
\STATE $D$ = $\{d \hspace{2mm}|\hspace{2mm} d = \arg\max P_i,|d| = K ,i \in \text{Best solution in the generation} \}$;
\STATE $S_N = \{N\} - D,S_s = \text{smaller set of selected assets}$;

\STATE $c = c_1$;
\IF{$rep \geq\alpha$}
\STATE $c = c_2$;
\IF{$rep \geq \beta $}
\STATE $r =\max(r+r,K-1)$;
\ENDIF
\ENDIF
\IF{$\text{random}(0,1)\leq c$ or $|S_s-D| \leq r$}
\STATE $I = D \bigcup \{j \hspace{2mm}|\hspace{2mm} j \in N - D,|j|=r \}$;
\ELSE
\STATE $I = D \bigcup \{j \hspace{2mm}|\hspace{2mm} j \in S_s - D,|j|=r \}$;
\ENDIF

\STATE Solve Model 5 with some specified time limit ($I = I,J=D$)\\

\end{algorithmic}
\end{algorithm}

Algorithm \ref{Our heuristic} explains GALB.

\begin{algorithm}[H]

\caption{GALB}
\begin{algorithmic} 
\label{Our heuristic}
\STATE Let $LB$ be the number of GA's iterations that has to pass to perform one Local branching;
\STATE Let $G=0,B=$ A large number; 
\STATE Initialize a random population $P$;
\WHILE{Condition has not met} 
\STATE $G=G+1$;
\STATE $C  = $ Crossover() ;
\STATE $M = $ Mutation() ;
\STATE $A = P \bigcup \{C \bigcup M \}$;
\STATE Sort $P_i$ and set the lowest $N-K$ weights to 0 and normalize it; $\forall i \in P$

\IF{$G = LB$}
\STATE Select the best solution and execute the Algorithm \ref{Local branching-alg};
\IF{The solution $\leq B$}
\STATE Add the solution to $A$;
\ENDIF
\STATE $G = 0$;
\ENDIF
\STATE $P=$ Best individuals of size $|P|$ chosen from $A$;
\STATE $b=$ Best solution in the generation $P$;
\IF{$b \leq B$}
\STATE $B = b$;
\ENDIF
\ENDWHILE

\end{algorithmic}
\end{algorithm}

Incorporating the local branching mechanism into GA helps exploit the combined capabilities
of both. On the one hand, GA is widely known for its global search capability, whereby it provides
a high-quality approximation in a short time. On the other hand, local branching is a local search 
technique that enhances the search process in GA. By performing local branching at specified 
intervals of GA iterations, we were able to strike a balance between the local and global search
processes.

\section{Experimental results:}
\label{Experimental results}
\begin{figure}[H]
\centering
\subfloat[CAC 40]{\includegraphics[width=0.49\textwidth]{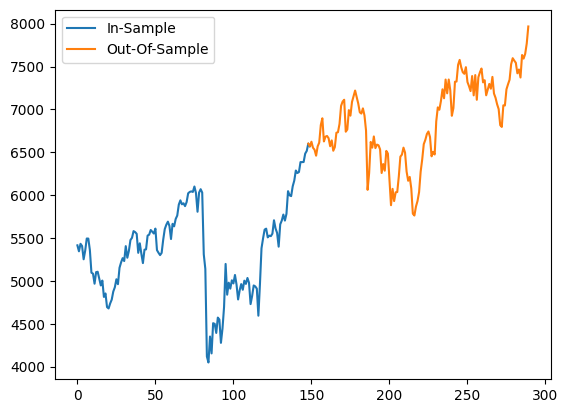}}
\hfill
\subfloat[EURO STOXX 50]{\includegraphics[width=0.49\textwidth]{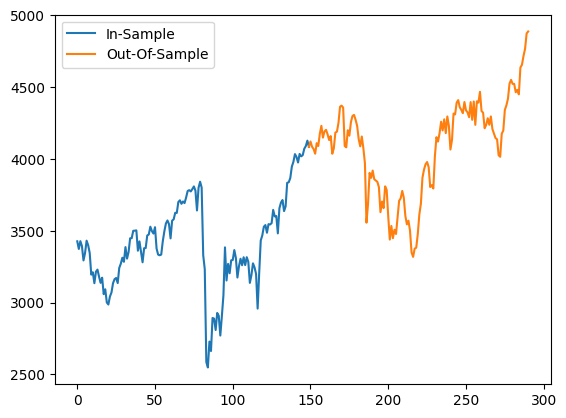}}
\hfill
\subfloat[Dow Jones]{\includegraphics[width=0.33\textwidth]{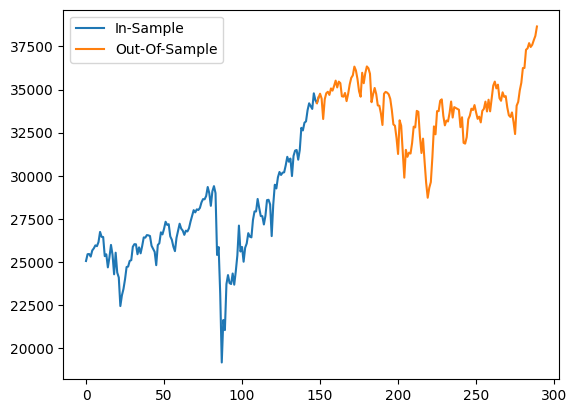}}
\hfill
\subfloat[Hang Seng]{\includegraphics[width=0.33\textwidth]{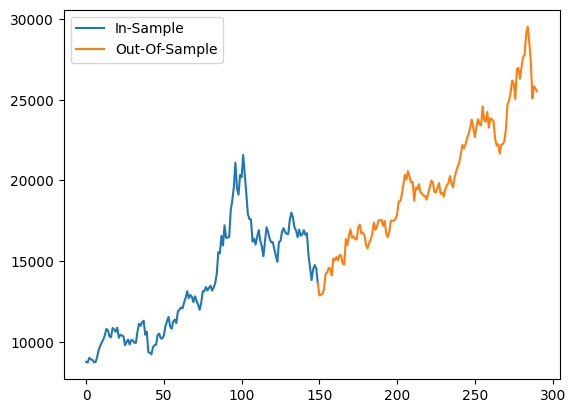}}
\hfill
\subfloat[DAX]{\includegraphics[width=0.33\textwidth]{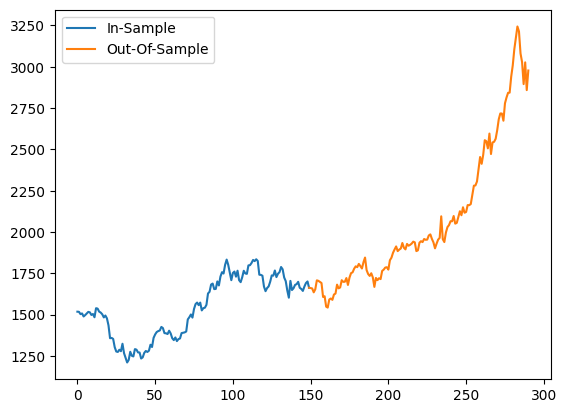}}
\hfill
\subfloat[FTSE 100]{\includegraphics[width=0.33\textwidth]{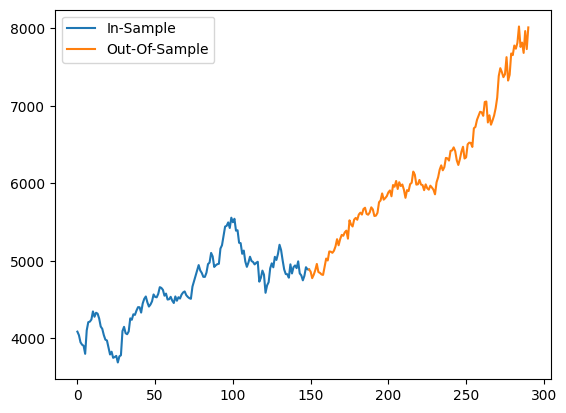}}
\hfill
\subfloat[S\&P100]{\includegraphics[width=0.33\textwidth]{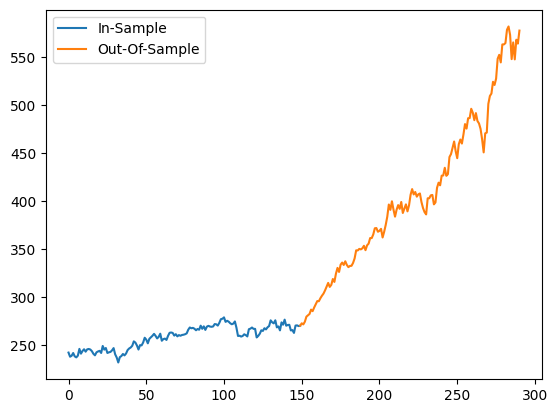}}
\hfill
\subfloat[nikkei 225]{\includegraphics[width=0.33\textwidth]{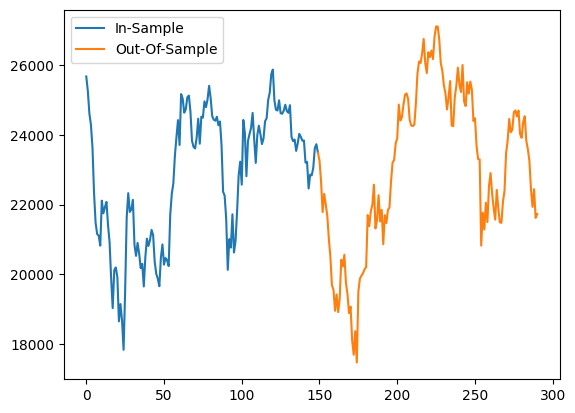}}
\caption{In-sample and our-of-sample periods for each market.}
\label{fig:data visual}
\end{figure}

This section compares the performance of the proposed heuristic approach with an exact 
solver. For this purpose, use was made of CPLEX as an exact solver for the benchmark. Moreover, 
the data from five well-known markets, namely, Hang Seng, DAX, FTSE 100, S\&P100 and nikkei 225 were extracted and used from OR-library. Furthermore, the data of the Dow Jones industrial 
average, CAC 40, and EURO STOXX 50 over the period from July 2018 to January 2024 
containing high, low, open, close, and volume of each asset, including the indices themselves, were 
downloaded from Yahoo Finance. Table \ref{tab:markets}  reports the number of assets in each market along with 
the standard deviations of returns of in-sample and out-of-sample periods. Figure \ref{fig:data visual} visualizes each 
index, distinguishing between in-sample and out-of-sample periods.

\begin{table}[H]
    \centering
    \caption{Overview of the markets}
    \label{tab:markets}
 \begin{tabular}{lccc}
\toprule
        Market & Number of assets & $\sigma$(I) & $\sigma$(O) \\
\midrule
        Dow Jones & 29 & 0.0329 & 0.0213 \\
        Hang Seng & 31 & 0.0377 & 0.0277 \\
        CAC 40    & 39 & 0.0329 & 0.0213 \\
        EURO STOXX 50 & 50 &0.0330&0.0237\\     
        DAX       & 85 & 0.0197 & 0.0208 \\
        FTSE 100  & 89 & 0.0192 & 0.0153 \\
        S\&P100   & 98 & 0.0124 & 0.0175 \\
        nikkei 225& 225& 0.0293 & 0.0279 \\
\bottomrule
    \end{tabular}
\end{table}

The rest of this Section is presented in two parts. The first deals with the design of an experiment 
to verify the effectiveness of our formulation as compared to some other commonly used ones. 
The second part is devoted to the design of another experiment that is used to evaluate the power 
of GALB. 

For the first experiment, the two formulations already introduced above, namely, Model 1 and 
Model 3, was considered for comparison with the one proposed here. As Model 1 required market 
capitalization data of assets, we were not able to compare our Model using OR-library data.
Instead, CAC 40, EURO STOXX 50, and Dow Jones were chosen for this purpose. Although it 
will be shown that our heuristic is capable of converging to optimal solutions in shorter times for
markets of the size mentioned before, we opted for the CPLEX solver in lieu of the proposed 
heuristic in order to completely eliminate the effect of randomness on the optimization process. In 
all the experiments presented in this paper, the values for $K$, $\epsilon$, and $\delta$ are set equal to 10, 1\%, and 
90\%, respectively. Furthermore, $\xi$ is assumed to be 100 in the proposed formulation. Finally, the 
following five metrics were considered for comparing the performance of the proposed 
formulation or GALB:

$\rho$: C Correlation between tracking portfolio’s returns and index returns.

\begin{flalign}\label{Eq:MAD}
\text{Mean Absolute Deviation (MAD)} = \dfrac{1}{T} \sum_{t=1}^T |I_t - \sum_{i=1}^N {r_{it}w_i}| &&
\end{flalign}

\begin{flalign}\label{Eq:RMSE}
\text{Root Mean Squared Error (RMSE)} = \sqrt{\dfrac{1}{T} \sum_{t=1}^T (I_t - \sum_{i=1}^N {r_{it}w_i})^2} &&
\end{flalign}

STD: Standard deviation of tracking errors.

Beta: Beta of tracking portfolio and index.

Even though smaller values of MAD, RMSE, and STD are desirable, larger values of $\rho$ and 
those of Beta close to unity are preferred. Tables \ref{tab:FC-DOW}, \ref{tab:FC-CAC}, and \ref{tab:FC-STOXX} below present comparisons of 
the formulations examined for Dow Jones, CAC 40, and EURO STOXX 50, respectively. Moreover, the in-sample results in these same Tables are marked by the letter (I) and the out-of-sample ones by the letter (O) in the parentheses placed before each metric. Model (\ref{Eq:weight1}) is represented in the Tables as Model 1 with the weighting method explained in equation (\ref{Eq:weight1}). Model 1 (J) is also the same as Model 1 except that the weight of asset $j$ in this model is the market cap of $j$ \footnote{Since Dow Jones is a price-weighted index, its weighting method in Model 1 is not as accurate as those of the other two indices.}. All the 
models are allowed 1800 seconds to generate a solution, while the values less than those reported 
in the relevant time row indicate the amount of time needed for CPLEX to yield an optimal 
solution. No time limit is assigned for Model 1 since it is a simple model, and CPLEX is capable 
of solving it almost instantly for the market sizes used in the present case.

\begin{table}[H]
    \centering
    \caption{Comparison of the formulations for Dow Jones industrial average(solved by CPLEX)}
    \label{tab:FC-DOW}
    \begin{tabular}{l>{\rowmac} c>{\rowmac} c>{\rowmac} c>{\rowmac} c<{\clearrow} }
        \toprule
     \multirow{1}{*}{} &
      \multicolumn{1}{c}{Our formulation($\Gamma = 0 $)} &
      \multicolumn{1}{c}{Model 1 (J)} &
      \multicolumn{1}{c}{Model 1 (\ref{Eq:weight1})} &
      \multicolumn{1}{c}{Model 3} \\
        \midrule

(I)$\rho$ &0.9897&0.9043 & 0.9361 & 0.9500 \\
(I)MAD  &0.00341&0.0100& 0.00880 &0.00324 \\
(I)RMSE &0.00503&0.01044&0.01174&0.00498 \\
(I)STD      &0.00491&0.01438&0.01160&0.00488 \\
(I)Beta     &1.0201&0.7386&0.8747&1.0180 \\
(I)Required time(s) &72&-&-&118 \\
(O)$\rho$ &0.9703&0.8664&0.8717&0.9604 \\
\setrow{\bfseries}(O)MAD &0.00402&0.00866&0.00872&0.00458\\
(O)RMSE &0.00539&0.01187&0.01186&0.00628\\
(O)STD      &0.00543&0.01188&0.01189&0.006321 \\
(O)Beta     &1.0177&0.9619&0.9898&1.0177\\
        \bottomrule
    \end{tabular}
\end{table}

\begin{table}[H]
    \centering
    \caption{Comparison of the formulations for CAC 40(solved by CPLEX)}
    \label{tab:FC-CAC}
    \begin{tabular}{l>{\rowmac} c>{\rowmac} c>{\rowmac} c>{\rowmac} c<{\clearrow} }
        \toprule
     \multirow{1}{*}{} &
      \multicolumn{1}{c}{Our formulation($\Gamma = 0 $)} &
      \multicolumn{1}{c}{Model 1(J)} &
      \multicolumn{1}{c}{Model 1(\ref{Eq:weight1})} &
      \multicolumn{1}{c}{Model 3} \\
        \midrule

(I)$\rho$ & 0.9928 & 0.9652 & 0.9369 & 0.9925 \\
(I)MAD  & 0.00315 & 0.00609 & 0.0112 & 0.00311 \\
(I)RMSE & 0.00430 & 0.00891 & 0.01554 & 0.00437 \\
(I)STD      & 0.00406 & 0.00871 & 0.01552 & 0.00420 \\
(I)Beta    & 1.0112 & 0.9497 & 1.1676 &  1.0173 \\
(I)Required time(s) &219&-&-&1800 \\
(O)$\rho$ & 0.9696 & 0.9349 & 0.9113 & 0.9604 \\
\setrow{\bfseries}(O)MAD  		& 0.00454 & 0.00706 & 0.00973 & 0.00458 \\
(O)RMSE & 0.00586 & 0.00915 & 0.01279 & 0.00628 \\
(O)STD      & 0.00576 & 0.00922 & 0.01282 & 0.00632 \\
(O)Beta     & 0.9148 & 1.032 & 1.1587 & 1.0177 \\
        \bottomrule
    \end{tabular}
\end{table}

A brief glance at the above Tables reveals that the portfolios produced by Model 1 exhibit poor 
results compared to those generated by Model 3 and ours.

As seen in Table \ref{tab:FC-DOW} CPLEX yielded an optimal solution for our formulation in 72 seconds but 
did the same for Model 3 in 118 seconds. Although the tracking portfolio produced from the 
proposed formulation demonstrated a stronger correlation with the market in the in-sample metric, 
it performed slightly worse with respect to the other metrics. When it comes to the out-of-sample 
results as is the main purpose of this study and IT endeavors, our formulation clearly outperforms
all the other models with respect to all the metrics. As regards CAC 40, similar to Dow Jones, our 
model’s portfolio exhibited better correlations and slightly higher MAD and RMSE values, as well 
as a better STD for in-sample results. In addition, CPLEX solved our formulation in 219 seconds, while it failed to provide an optimal solution for Model 3 in 1800 seconds. Similar to Dow Jones,
the out-of-sample results generated by the proposed model outperformed all the other formulations
examined in terms of all the metrics except for Beta. For the last market, CPLEX failed to obtain 
an optimal solution within 1800 seconds, and further, it generated identical solutions for both 
Model 3 and the one proposed in this paper. Moreover, Model 1 portfolios are still the worst 
possible portfolios that can be chosen.

\begin{table}[H]
    \centering
    \caption{Comparison of formulations for EURO STOXX 50(solved by CPLEX)}
    \label{tab:FC-STOXX}
    \begin{tabular}{l>{\rowmac} c>{\rowmac} c>{\rowmac} c>{\rowmac} c<{\clearrow} }
        \toprule
     \multirow{1}{*}{} &
      \multicolumn{1}{c}{Our formulation($\Gamma = 0 $)} &
      \multicolumn{1}{c}{Model 1 (J)} &
      \multicolumn{1}{c}{Model 1 (\ref{Eq:weight1})} &
      \multicolumn{1}{c}{Model 3} \\
        \midrule

(I)$\rho$ & 0.9901 & 0.4587 & 0.6966 & 0.9901 \\
(I)MAD  & 0.00355 & 0.0370 & 0.0225 & 0.00355 \\
(I)RMSE & 0.00521 & 0.04951 & 0.03002 & 0.00521 \\
(I)STD      & 0.00486 & 0.04916 & 0.02951 & 0.00486 \\
(I)Beta    & 1.0243 & 0.7554 & 0.8529 & 1.0243 \\
(I)Required time(s) &1800&-&-&1800 \\
(O)$\rho$ & 0.9531 & 0.5781 & 0.7239 & 0.9531 \\
\setrow{\bfseries}(O)MAD 		& 0.00571 & 0.00706 & 0.03766 & 0.00571 \\
(O)RMSE & 0.00728 & 0.06048 & 0.03677 & 0.00728 \\
(O)STD      & 0.00732 & 0.06089 & 0.03702 & 0.00732 \\
(O)Beta     & 0.9594 & 1.7342 & 1.5321 & 0.9594 \\
        \bottomrule
    \end{tabular}
\end{table}

The second part of the experiments was meant to test the performance of GALB against the 
CPLEX solver. The tests were performed using four different $\Gamma$ values (namely, $\Gamma=0$ 0, indicating
no uncertainty included, as well as $\Gamma=50\%$,$\Gamma=75\%$ and $\Gamma=90\%$ indicating that 50\%,75\% and 90\% respectively, of the parameters took their worst possible values). Moreover, periods 
of 10 weeks were considered for calculating downside deviations. To provide a fair comparison 
between the two mechanisms examined, the time limit for the CPLEX solver was set to 1800 seconds 
for all the markets, while that for the proposed heuristic was set to 120 seconds for Dow Jones and 
Hang seng but 300 seconds for the other markets. Additionally, two new performance metrics were
added for the in-sample; these included the best Objective Function Value (OFV) and standard 
deviation of OFV (OFV($\sigma$)), defined as the standard deviation of five runs. Since the problem is 
one of minimization, the lower values of both metrics are desirable. Just as in the case of the 
formulation tests, there are both in-sample and out-of-sample results. For the in-sample values,
lower OFV values are expected to be fully correlated with the in-sample metrics, meaning that 
lower OFVsshould lead to better in-sample results and vice versa. However, as claimed by (\cite{beasley2003evolutionary}) even though in-sample and out-of-sample results are correlated, good in-sample results 
do not necessarily entail good out-of-sample ones. Nevertheless, superior in-sample results 
expectedly lead to better out-of-sample ones.

Table \ref{tab:Dow-10} reports the results obtained for the Dow Jones market. It is obvious from the OFV values
that our heuristic outperformed the CPLEX solver in all $\Gamma$ settings, although the results were quite 
competitive. Moreover, the heuristic converged to global optima in each run for $\Gamma = 0$, and both 
the CPLEX and the heuristic yielded identical portfolios. Furthermore, the best in-sample and out-of-sample results were achieved by our heuristic when $\Gamma = 90 \%$. Table \ref{tab:hang-seng10} reports the results for 
the Hang Seng market. Similar to its performance in the case of Dow Jones, the proposed heuristic 
not only converged to the optimal solution for the first setup in each run but also outperformed
CPLEX in all the metrics with respect to the in-sample results under all the setups except for two 
setups in beta (namely, $\Gamma = 50\%, \Gamma = 75\%$).  Furthermore, disregarding Beta, the heuristic also 
outperformed CPLEX with respect to all the out-of-sample metrics, while the best results were 
obtained for $\Gamma = 75\%$.

\begin{table}[H]
    \centering
    \caption{Dow Jones Industrial Average, CPLEX = 1800s, GALB = 120s}
    \label{tab:Dow-10}
    \begin{tabular}{l>{\rowmac}c>{\rowmac}c>{\rowmac}c>{\rowmac}c>{\rowmac}c>{\rowmac} c>{\rowmac}c>{\rowmac}c<{\clearrow}}
        \toprule

 
     \multirow{1}{*}{} &
      \multicolumn{1}{c}{CPELX} &
      \multicolumn{1}{c}{GALB} &
      \multicolumn{1}{c}{CPELX} &
      \multicolumn{1}{c}{GALB} &
      \multicolumn{1}{c}{CPELX} &
      \multicolumn{1}{c}{GALB} &
      \multicolumn{1}{c}{CPELX} &
      \multicolumn{1}{c}{GALB}\\
      &  $\Gamma = 0$ & $\Gamma = 0$ & $\Gamma = 50\%$ & $\Gamma = 50\%$ & $\Gamma = 75\%$ & $\Gamma = 75\%$ & $\Gamma = 90\%$ & $\Gamma = 90\%$ \\
        \midrule
(I)OFV        & -0.6209 & -0.6209 & -0.4417 & -0.4445 & -0.4033 & -0.4078 & -0.4021 & -0.4078 \\
(I)OFV($\sigma$)& - & 0 & - & 0.0087 & - & 0.0120 & - & 0.0077 \\
(I)$\rho$   & 0.9897 & 0.9897 & 0.9893 & 0.9895 & 0.9893 & 0.9898 & 0.9896 & 0.9918 \\
(I)MAD    & 0.00341 & 0.00341 & 0.00335 & 0.00336 & 0.00335 & 0.00336 & 0.00328 & 0.00329 \\
(I)RMSE   & 0.00503 & 0.00503 & 0.00492 & 0.00488 & 0.00492 & 0.00489 & 0.00490 & 0.00468  \\
(I)STD        & 0.00491 & 0.00491 & 0.00483 & 0.00480 & 0.00483 & 0.00480 & 0.00483 & 0.00437  \\
(I)Beta       & 1.0201 & 1.0201 & 0.9942 & 0.9967 & 0.9942 & 0.9967 & 1.0079 & 1.0204  \\
(O)$\rho$   & 0.9703 & 0.9703 & 0.9626 & 0.9653 & 0.9626 & 0.9653 & 0.9633 & 0.9752 \\
\setrow{\bfseries}(O)MAD   & 0.00402 & 0.00402 & 0.00444 & 0.00424 & 0.00444 & 0.00424 & 0.00439 & 0.00388\\
(O)RMSE   & 0.00539 & 0.00539 & 0.00599 & 0.00574 & 0.00599 & 0.00574 & 0.00604 & 0.00497\\
(O)STD        & 0.00543 & 0.00543 & 0.00603 & 0.00578 & 0.00603 & 0.00578 & 0.00609 & 0.00491\\
(O)Beta       & 1.0177 & 1.0177 & 1.0044 & 1.0017 & 1.0044 & 1.0017 & 1.0203 & 1.0127\\

        \bottomrule
    \end{tabular}
\end{table}

\begin{table}[H]
    \centering
    \caption{Hang Seng, CPLEX = 1800s, GALB = 120s}
    \label{tab:hang-seng10}
    \begin{tabular}{l>{\rowmac}c>{\rowmac}c>{\rowmac}c>{\rowmac}c>{\rowmac}c>{\rowmac} c>{\rowmac}c>{\rowmac}c<{\clearrow}}
        \toprule

 
     \multirow{1}{*}{} &
      \multicolumn{1}{c}{CPELX} &
      \multicolumn{1}{c}{GALB} &
      \multicolumn{1}{c}{CPELX} &
      \multicolumn{1}{c}{GALB} &
      \multicolumn{1}{c}{CPELX} &
      \multicolumn{1}{c}{GALB} &
      \multicolumn{1}{c}{CPELX} &
      \multicolumn{1}{c}{GALB}\\
      &  $\Gamma = 0$ & $\Gamma = 0$ & $\Gamma = 50\%$ & $\Gamma = 50\%$ & $\Gamma = 75\%$ & $\Gamma = 75\%$ & $\Gamma = 90\%$ & $\Gamma = 90\%$ \\
        \midrule
(I)OFV & -0.6597 & -0.6597 & -0.5297 & -0.5306 & -0.5030 & -0.5068 & -0.5030 & -0.5096 \\

(I)OFV($\sigma$)& - & 0 & - & 0.0099 & - & 0.0044 & - & 0.0097 \\


(I)$\rho$ & 0.9951 & 0.9951 & 0.9941 & 0.9949 & 0.9941 & 0.9948 & 0.9941 & 0.9951 \\

(I)MAD    & 0.00281 & 0.00281 & 0.00309 & 0.00287 & 0.00309 & 0.00286 & 0.00309 & 0.00281 \\
(I)RMSE  & 0.00373 & 0.00373 & 0.00406 & 0.00383 & 0.00406 & 0.00384 & 0.00406 & 0.00373 \\
(I)STD       & 0.00373 & 0.00373 & 0.00407 & 0.00382 & 0.00407 & 0.00384 & 0.00407 & 0.00373 \\
(I)Beta      & 0.9955 & 0.9955 & 0.9918 & 0.9783 & 0.9918 & 0.9894 & 0.9918 & 0.9955 \\
(O)$\rho$ & 0.9862 & 0.9862 & 0.9818 & 0.9855 & 0.9818 & 0.9866 & 0.9818 & 0.9862 \\
\setrow{\bfseries}(O)MAD  & 0.00352 & 0.00352 & 0.00389 & 0.00340 & 0.00389 & 0.00337 & 0.00389 & 0.00352\\
(O)RMSE  & 0.00468 & 0.00468 & 0.00526 & 0.00464 & 0.00526 & 0.00449 & 0.00526 & 0.00468\\
(O)STD      & 0.00471 & 0.00471 & 0.00530 & 0.00465 & 0.00530 & 0.00452 & 0.00530 & 0.00471\\
(O)Beta     & 1.0202 & 1.0202 & 1.0016 & 0.9845 & 1.0016 & 0.9995 & 1.0016 & 1.0202\\

        \bottomrule
    \end{tabular}
\end{table}

Table \ref{tab:DAX-10} provides comparisons among the different approaches for the case of the DAX market. Clearly, the portfolio generated by our approach exhibited a better performance in all the out-of-sample metrics except for $\Gamma = 50\%$, while CPLEX showed a better performance in the first setting on in-sample data. 
It is also seen that the differences in the OFV values between CPLEX and GALB increased compared
with those of the markets under other setups, with the best out-of-sample tracking portfolio
obtained with the setting GALB $\Gamma = 75\%$. Even though our heuristic achieved better in-sample results
with $\Gamma = 50\%$, this was not true for out-of-sample ones as the CPLEX portfolio showed a superior
performance in this case. Moreover, the gap between OFVs increased with increasing $\Gamma$.

The comparisons between CPLEX and our approach are reported in Table \ref{tab:FTSE-10} for FTSE 100. This
is the first market for which not only did our heuristic find better OFV values even under the first 
setting but also the first and the only one for which the CPLEX solver’s portfolio yielded the best 
out-of-sample results (See the out-of-sample metrics for CPLEX, $\Gamma = 50\%$) even in the face of
the fact that our heuristic yielded much better in-sample results in this case. Besides, our heuristic 
outperformed the CPLEX solver with respect to both in-sample and out-of-sample data.
 
 \begin{table}[H]
    \centering
    \caption{DAX, CPLEX = 1800s, GALB = 300s}
    \label{tab:DAX-10}
    \begin{tabular}{l>{\rowmac}c>{\rowmac}c>{\rowmac}c>{\rowmac}c>{\rowmac}c>{\rowmac} c>{\rowmac}c>{\rowmac}c<{\clearrow}}
        \toprule

 
     \multirow{1}{*}{} &
      \multicolumn{1}{c}{CPELX} &
      \multicolumn{1}{c}{GALB} &
      \multicolumn{1}{c}{CPELX} &
      \multicolumn{1}{c}{GALB} &
      \multicolumn{1}{c}{CPELX} &
      \multicolumn{1}{c}{GALB} &
      \multicolumn{1}{c}{CPELX} &
      \multicolumn{1}{c}{GALB}\\
      &  $\Gamma = 0$ & $\Gamma = 0$ & $\Gamma = 50\%$ & $\Gamma = 50\%$ & $\Gamma = 75\%$ & $\Gamma = 75\%$ & $\Gamma = 90\%$ & $\Gamma = 90\%$ \\
        \midrule
(I)OFV        & -0.7385 & -0.7377 & -0.4202 & -0.4329 & -0.2686 & -0.3118 & -0.2171 & -0.2723 \\
(I)OFV($\sigma$)& - & 0.0010 & - & 0.0059 & - & 0.0052 & - & 0.0029 \\
(I)$\rho$   & 0.9881 & 0.9867 & 0.9894 & 0.9892 & 0.9883 & 0.9881 & 0.9851 & 0.9879 \\
(I)MAD    & 0.00233 & 0.00251 & 0.00233 & 0.00217 & 0.00260 & 0.00236 & 0.00233 & 0.00237 \\
(I)RMSE  & 0.00320 & 0.00336 & 0.00306 & 0.00294 & 0.00337 & 0.00317 & 0.00320 & 0.00317  \\
(I)STD        & 0.00322 & 0.00338 & 0.003081 & 0.00294 & 0.00338 & 0.00319 & 0.00322 & 0.00319  \\
(I)Beta       & 1.0303 & 1.0276 & 1.0351 & 1.0049 & 1.0540, & 1.0253 & 1.0303 & 1.0226  \\
(O)$\rho$   & 0.8984 & 0.9001 & 0.9119 & 0.9077 & 0.9052 & 0.9116 & 0.8984 & 0.9063 \\
\setrow{\bfseries}(O)MAD    & 0.00496 & 0.00488 & 0.00440 & 0.00470 & 0.00479 & 0.00440 & 0.00496 & 0.00474\\
(O)RMSE   & 0.00926 & 0.00918 & 0.00862 & 0.00877 & 0.00892 & 0.00863 & 0.00926 & 0.00882\\
(O)STD        & 0.00932 & 0.00924 & 0.00868 & 0.00883 & 0.00898 & 0.00869 & 0.00932 & 0.00889\\
(O)Beta       & 0.8759 & 0.8762 & 0.8859 & 0.8653 & 0.8734 & 0.8856 & 0.8759 & 0.8612\\

        \bottomrule
    \end{tabular}
\end{table}
\begin{table}[H]
    \centering
    \caption{FTSE 100, CPLEX = 1800s, GALB = 300s}
    \label{tab:FTSE-10}
    \begin{tabular}{l>{\rowmac}c>{\rowmac}c>{\rowmac}c>{\rowmac}c>{\rowmac}c>{\rowmac} c>{\rowmac}c>{\rowmac}c<{\clearrow}}
        \toprule

 
     \multirow{1}{*}{} &
      \multicolumn{1}{c}{CPELX} &
      \multicolumn{1}{c}{GALB} &
      \multicolumn{1}{c}{CPELX} &
      \multicolumn{1}{c}{GALB} &
      \multicolumn{1}{c}{CPELX} &
      \multicolumn{1}{c}{GALB} &
      \multicolumn{1}{c}{CPELX} &
      \multicolumn{1}{c}{GALB}\\
      &  $\Gamma = 0$ & $\Gamma = 0$ & $\Gamma = 50\%$ & $\Gamma = 50\%$ & $\Gamma = 75\%$ & $\Gamma = 75\%$ & $\Gamma = 90\%$ & $\Gamma = 90\%$ \\
        \midrule
(I)OFV        & -0.5542 & -0.5639 & -0.1684 & -0.2317 & +0.0170 & -0.1366 & -0.0293 & -0.0312 \\
(I)OFV($\sigma$)& - & 0.0048 & - & 0.0164 & - & 0.3146 & - & 0.0111 \\
(I)$\rho$   & 0.9659 & 0.9630 & 0.9427 & 0.9566 & 0.9259 & 0.9499 & 0.9566 & 0.9452 \\
(I)MAD    & 0.00410 & 0.00406 & 0.00504 & 0.00419 & 0.00580 & 0.00416 & 0.00487 & 0.00475 \\
(I)RMSE   & 0.00570 & 0.00550 & 0.00682 & 0.00579 & 0.00760 & 0.00571 & 0.00646 & 0.00646  \\
(I)STD        & 0.00567 & 0.00550 & 0.006860 & 0.00582 & 0.00763 & 0.00572 & 0.00646 & 0.00649  \\
(I)Beta       & 1.0670 & 1.0165 & 1.003 & 0.9889 & 0.9650 & 0.9665 & 1.0079 & 0.9689  \\
(O)$\rho$   & 0.8672 & 0.8976 & 0.9101 & 0.9091 & 0.8485 & 0.8781 & 0.8686 & 0.8775 \\
\setrow{\bfseries}(O)MAD    & 0.00614 & 0.00564 & 0.00493 & 0.00511 & 0.00687 & 0.00594 & 0.00638 & 0.00591\\
(O)RMSE   & 0.00793 & 0.00698 & 0.00647 & 0.00656 & 0.00859 & 0.00750 & 0.00789 & 0.00763\\
(O)STD        & 0.00798 & 0.00703 & 0.00651 & 0.00651 & 0.00859 & 0.00762 & 0.00794 & 0.00759\\
(O)Beta       & 0.8835 & 0.9188 & 0.9125 & 0.9018 & 0.8763 & 0.8901 & 0.8861 & 0.8784\\

        \bottomrule
    \end{tabular}
\end{table}

Table \ref{tab:SP-10} compares the results obtained for the S\&P 100 market. In the first setting, the proposed 
heuristic and CPLEX yielded identical portfolios. Under $\Gamma = 50\%$, CPLEX yielded more 
promising out-of-sample results even though more favorable in-sample performance were 
obtained with the proposed heuristic. In the other two cases, the heuristic outperformed CPLEX 
(with a high difference margin in the out-of-sample data). Last but not least, a comparison of both 
approaches in nikkei 225 revealed the superiority of the proposed heuristic in all the metrics
studied over the CPLEX solver with regard to in-sample and out-of-sample data, except for $\Gamma = 75\%$ that the heuristic did show a better in-sample performance but failed to yield better out-of-sample results.
\begin{table}[H]
    \centering
    \caption{S\&P 100, CPLEX = 1800s, GALB = 300s}
    \label{tab:SP-10}
    \begin{tabular}{l>{\rowmac}c>{\rowmac}c>{\rowmac}c>{\rowmac}c>{\rowmac}c>{\rowmac} c>{\rowmac}c>{\rowmac}c<{\clearrow}}
        \toprule

 
     \multirow{1}{*}{} &
      \multicolumn{1}{c}{CPELX} &
      \multicolumn{1}{c}{GALB} &
      \multicolumn{1}{c}{CPELX} &
      \multicolumn{1}{c}{GALB} &
      \multicolumn{1}{c}{CPELX} &
      \multicolumn{1}{c}{GALB} &
      \multicolumn{1}{c}{CPELX} &
      \multicolumn{1}{c}{GALB}\\
      &  $\Gamma = 0$ & $\Gamma = 0$ & $\Gamma = 50\%$ & $\Gamma = 50\%$ & $\Gamma = 75\%$ & $\Gamma = 75\%$ & $\Gamma = 90\%$ & $\Gamma = 90\%$ \\
        \midrule
(I)OFV        & -0.6121 & -0.6121 & -0.1986 & -0.2323 & -0.0067 & -0.0917 & +0.09346 & -0.0181 \\
(I)OFV($\sigma$)& - & 0.0058 & - & 0.0121 & - & 0.0183 & - & 0.0203 \\
(I)$\rho$   & 0.9282 & 0.9282 & 0.9307 & 0.9376 & 0.9066 & 0.9408 & 0.9151 & 0.9299 \\
(I)MAD    & 0.00357 & 0.00357 & 0.00384 & 0.00365 & 0.00434 & 0.00339 & 0.00428 & 0.00353 \\
(I)RMSE   & 0.00490 & 0.00490 & 0.00489 & 0.00487 & 0.00598 & 0.00443 & 0.00569 & 0.00481  \\
(I)STD        & 0.00492 & 0.00492 & 0.00486 & 0.00485 & 0.00602 & 0.00439 & 0.00570 & 0.00479  \\
(I)Beta       & 0.9951 & 0.9951 & 1.0023 & 1.0508 & 1.0436 & 0.9881 & 1.0438 & 0.9806  \\
(O)$\rho$   & 0.9311 & 0.9311 & 0.9341 & 0.9155 & 0.9174 & 0.9313 & 0.9019 & 0.9479 \\
\setrow{\bfseries}(O)MAD    & 0.00559 & 0.00559 & 0.00533 & 0.00590 & 0.00669 & 0.00511 & 0.00609 & 0.00454\\
(O)RMSE   & 0.00692 & 0.00692 & 0.00660 & 0.00743 & 0.00816 & 0.00656 & 0.00781 & 0.00583\\
(O)STD        & 0.00696 & 0.00696 & 0.00665 & 0.00748 & 0.00819 & 0.00661 & 0.00787 & 0.00585\\
(O)Beta       & 1.0119 & 1.0119 & 0.9907 & 0.9672 & 1.0650 & 0.9564 & 0.9218 & 0.9914\\
        \bottomrule
    \end{tabular}
\end{table}

\begin{table}[H]
    \centering
    \caption{nikkei 225,  CPLEX = 1800s, GALB = 300s}
    \label{tab:nikkei-10}
    \begin{tabular}{l>{\rowmac}c>{\rowmac}c>{\rowmac}c>{\rowmac}c>{\rowmac}c>{\rowmac} c>{\rowmac}c>{\rowmac}c<{\clearrow}}
        \toprule

 
     \multirow{1}{*}{} &
      \multicolumn{1}{c}{CPELX} &
      \multicolumn{1}{c}{GALB} &
      \multicolumn{1}{c}{CPELX} &
      \multicolumn{1}{c}{GALB} &
      \multicolumn{1}{c}{CPELX} &
      \multicolumn{1}{c}{GALB} &
      \multicolumn{1}{c}{CPELX} &
      \multicolumn{1}{c}{GALB}\\
      &  $\Gamma = 0$ & $\Gamma = 0$ & $\Gamma = 50\%$ & $\Gamma = 50\%$ & $\Gamma = 75\%$ & $\Gamma = 75\%$ & $\Gamma = 90\%$ & $\Gamma = 90\%$ \\
        \midrule
(I)OFV        & -0.5526 & -0.5634 & -0.1697 & -0.3216 & -0.1016 & -0.2126 & -0.0877 & -0.1869 \\
(I)OFV($\sigma$)& - & 0.0255 & - & 0.0355 & - & 0.0056 & - & 0.0399 \\
(I)$\rho$   & 0.9811 & 0.9830 & 0.9692 & 0.9774 & 0.9721 & 0.9713 & 0.9717 & 0.9789 \\
(I)MAD    & 0.00414 & 0.00401 & 0.00543 & 0.00485 & 0.00530 & 0.00507 & 0.00532 & 0.00444 \\
(I)RMSE   & 0.00571 & 0.00544 & 0.00749 & 0.00617 & 0.00724 & 0.00708 & 0.00703 & 0.00635  \\
(I)STD        & 0.00572 & 0.00546 & 0.00748 & 0.00621 & 0.00722 & 0.00710 & 0.00703 & 0.00636  \\
(I)Beta       & 0.9868 & 0.9977 & 1.0028 & 0.9672 & 1.0162 & 0.9845 & 0.9826 & 1.0298  \\
(O)$\rho$   & 0.9442 & 0.9686 & 0.9432 & 0.9641 & 0.9567 & 0.9514 & 0.9528 & 0.9641 \\
\setrow{\bfseries}(O)MAD    & 0.00688 & 0.00618 & 0.00720 & 0.00596 & 0.00656 & 0.00678 & 0.00666 & 0.00587\\
(O)RMSE   & 0.00963 & 0.00783 & 0.00926 & 0.00768 & 0.00833 & 0.00898 & 0.00886 & 0.00775\\
(O)STD        & 0.00940 & 0.00788 & 0.00931 & 0.00773 & 0.00838 & 0.00904 & 0.00890 & 0.00779\\
(O)Beta       & 0.9542 & 1.0671 & 0.9103 & 1.004 & 0.9834 & 0.9985 & 0.9991 & 1.0103\\

        \bottomrule
    \end{tabular}
\end{table}

\begin{figure}[H]
\centering
\subfloat[Dow-Jones]{\includegraphics[width=0.33\textwidth]{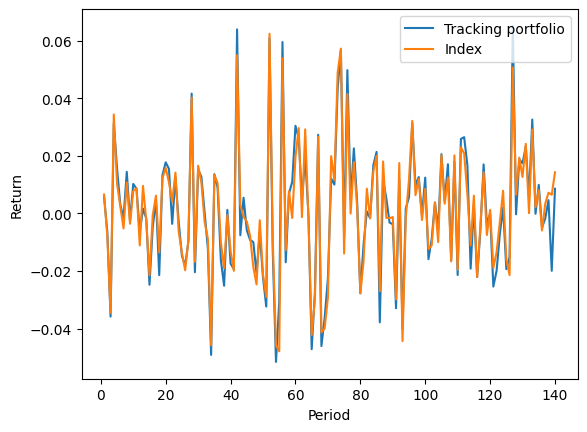}}
\hfill
\subfloat[Hang Seng]{\includegraphics[width=0.33\textwidth]{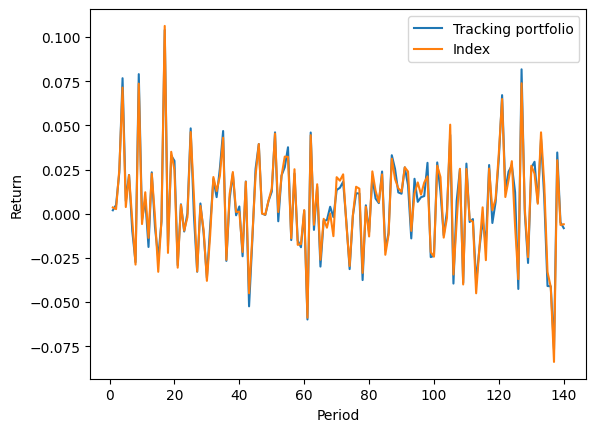}}
\hfill
\subfloat[DAX]{\includegraphics[width=0.33\textwidth]{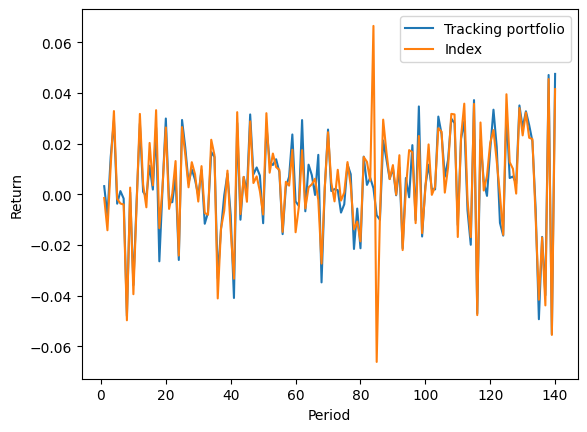}}
\hfill
\subfloat[FTSE 100]{\includegraphics[width=0.33\textwidth]{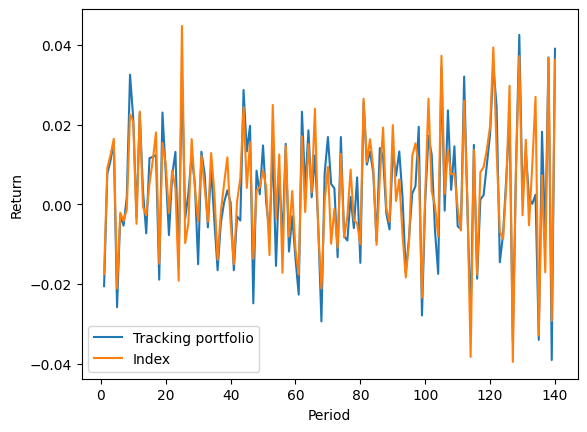}}
\hfill
\subfloat[S\&P 100]{\includegraphics[width=0.33\textwidth]{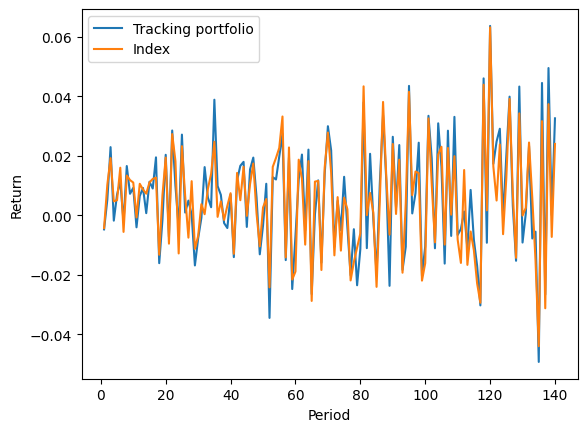}}
\hfill
\subfloat[nikkei 225]{\includegraphics[width=0.33\textwidth]{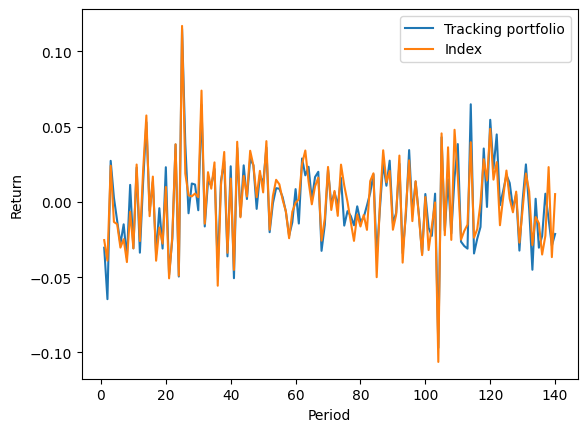}}
\hfill
\caption{Out-of-sample returns produced by our heuristic.}
\label{fig:returns}
\end{figure}

\begin{figure}[H]
\centering
\subfloat[Dow-Jones]{\includegraphics[width=0.33\textwidth]{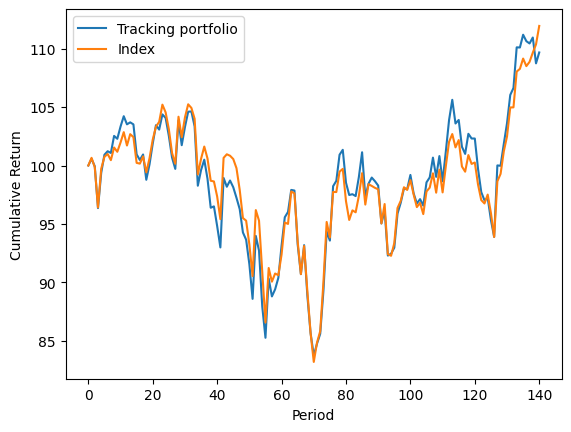}}
\hfill
\subfloat[Hang Seng]{\includegraphics[width=0.33\textwidth]{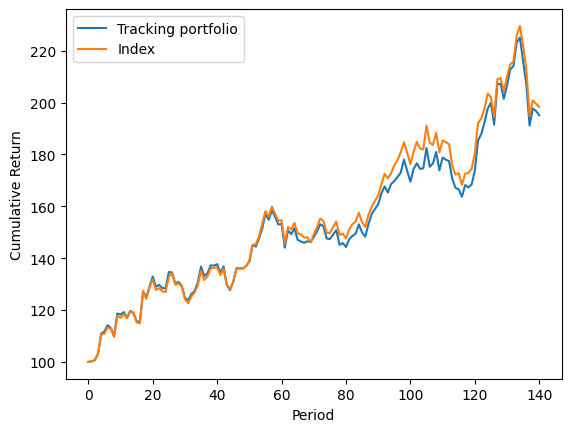}}
\hfill
\subfloat[DAX]{\includegraphics[width=0.33\textwidth]{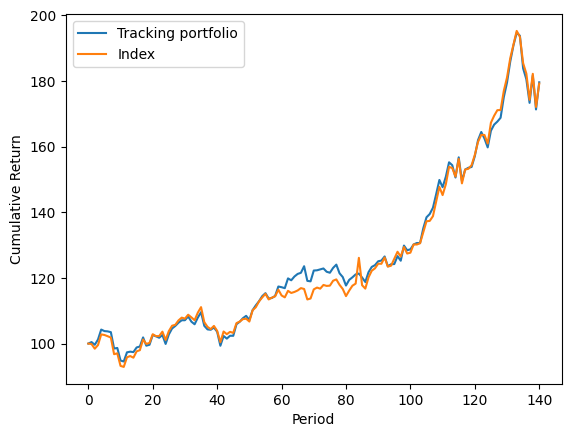}}
\hfill
\subfloat[FTSE 100]{\includegraphics[width=0.33\textwidth]{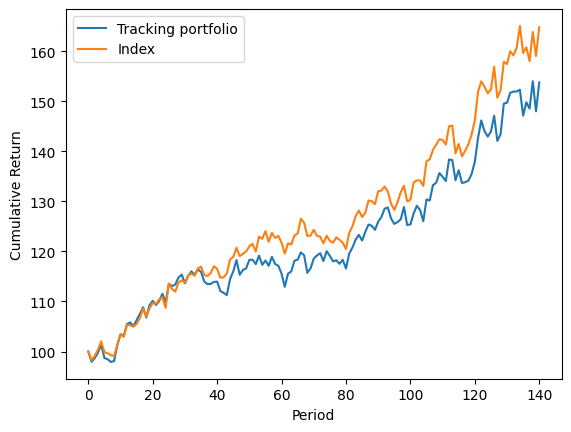}}
\hfill
\subfloat[S\&P 100]{\includegraphics[width=0.33\textwidth]{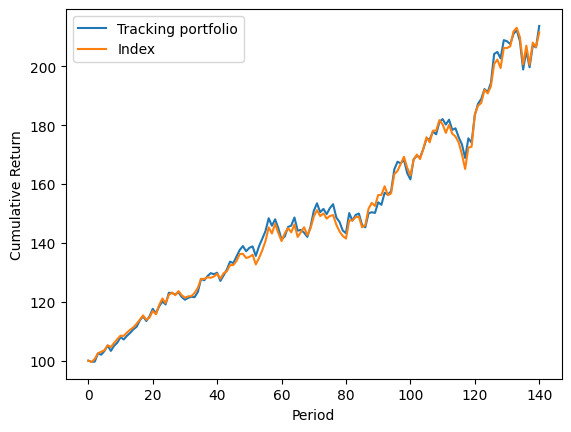}}
\hfill
\subfloat[nikkei 225]{\includegraphics[width=0.33\textwidth]{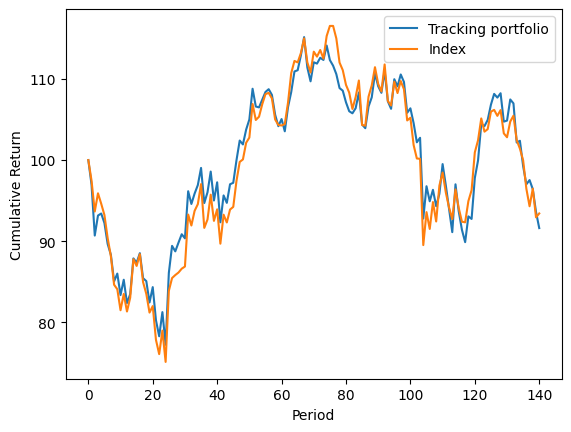}}
\hfill
\caption{Out-of-sample cumulative returns produced by our heuristic.}
\label{fig:cumsum}
\end{figure}

Figure \ref{fig:returns} and \ref{fig:cumsum} illustrate both out-of-sample and cumulative returns, respectively, of the 
tracking portfolios generated by our heuristic for each of the indices cited. The last Figure indicates
what happens to both the index and the tracking portfolio generated by the heuristic at the 
beginning of the period if a fixed amount of cash is invested. Evidently, the portfolios yielded by 
the proposed heuristic show outstanding results and extremely close tracking indices even in noisy 
and volatile markets. In addition, tracking portfolios and indices exhibit the same trend over long 
periods of time without much deviation or the need for rebalancing, which can reduce the fund 
managing costs. Finally, it is fair to remember that all portfolios are composed of 10 assets. Such portfolios are tracking markets with a large universe of assets 
such as nikkei 225 and S\&P 100 with such low deviations with no need for rebalancing.

\section{Conclusion:}
\label{Conclusion}
A novel and robust mixed-integer linear programming mathematical model was proposed for 
the index tracking problem. The proposed model was compared with two previously reported
formulations using the CPLEX solver and run on Dow Jones, CAC 40, and EURO STOXX 50. 
Results indicated that CPLEX solved our model faster and that the portfolios generated based on
our formulation using CPLEX yielded better out-of-sample data. Moreover, a novel heuristic 
(GALB) was proposed to solve the problem for increasing problem sizes and compared it with a 
commercial solver (CPLEX) using the benchmark data of OR-library. Results showed that GALB 
was found capable of not only finding better solutions in shorter times but also converging to the 
optimal solution for not-too-large problem sizes. Analysis of the results showed that the propsoed
approach was able to generate tracking portfolios that were much more correlated with the index
considered, exhibiting low values of the tracking error measures used (such as mean absolute 
deviation, root mean squared error, and standard deviation of tracking error). Finally, the generated 
portfolios that comprisd asset numbers of as low as 10 were found capable of tracking indices over
long periods of time without any considerable deviaitions from the index or the need for 
rebalancing even in larger and volatile markets.

\section*{Conflict of interest:}
The authors declare no conflict of interest.
\section*{Ethics approval:}
Not applicable.
\section*{Funding:}
This research received no specific grant from any funding agency in the public, commercial, or not-for-profit sectors.
\section*{Data availability:}
The data used in this study are publicly available and referenced in the text.
\section*{Authors contribution:}

\noindent\textbf{Danial Ramezani:} Conceptualization, Methodology, Software, Writing- Original draft preparation.

\noindent\textbf{Mostafa Abouei Ardakan:} Supervision, Conceptualization, Methodology, Writing- Reviewing and Editing.

\noindent\textbf{Mohamadreza Dehghani Ahmadabad:} Supervision, Conceptualization, Methodology, Writing- Reviewing and Editing.
\section*{Acknowledgement:}
No acknowledgments to declare.
\section*{Human participants and/or animals:}
Not applicable.

\renewcommand{\bibname}{References} %
\makeatletter
\renewcommand{\@biblabel}[1]{#1.} 
\renewcommand{\bibsection}{%
  \section*{\bibname}
  \addcontentsline{toc}{section}{\bibname}
}
\makeatother

\bibliographystyle{plainnat}
\bibliography{index-ref}

@article{silva2023systematic,
  title={A systematic literature review on solution approaches for the index tracking problem in the last decade},
  author={Silva, Julio Cezar Soares and others},
  journal={arXiv preprint arXiv:2306.01660},
  year={2023}
}

@article{yang2023new,
  title={A new uncertain enhanced index tracking model with higher-order moment of the downside},
  author={Yang, Tingting and Huang, Xiaoxia and Hong, Kwon Ryong},
  journal={Soft Computing},
  volume={27},
  number={16},
  pages={11379--11394},
  year={2023},
  publisher={Springer}
}

@article{sammon2026index,
  title={Index rebalancing and stock market composition: Do indexes time the market?},
  author={Sammon, Marco and Shim, John J},
  journal={Journal of Financial Economics},
  volume={177},
  pages={104229},
  year={2026},
  publisher={Elsevier}
}

@article{wang2026exact,
  title={An exact algorithm for a cardinality-constrained index tracking model considering investment preferences in portfolio optimization},
  author={Wang, Chun and Wu, Zhongming and Xu, Wei and Yuan, Yu},
  journal={Journal of Industrial and Management Optimization},
  volume={22},
  number={1},
  pages={612--641},
  year={2026}
}

@article{cesarone2025benchmark,
  title={A benchmark-asset principal component factorization for index tracking on large investment universes},
  author={Cesarone, Francesco and Di Paolo, Alessio and Bufalo, Michele and Orlando, Giuseppe},
  journal={Finance Research Letters},
  volume={79},
  pages={107244},
  year={2025},
  publisher={Elsevier}
}

@article{fieberg2025enhancing,
  title={Enhancing index-tracking performance: Leveraging characteristic-based factor models for reduced estimation errors},
  author={Fieberg, Christian and Osorio, Carlos and Poddig, Thorsten and Varmaz, Armin},
  journal={European Journal of Operational Research},
  year={2025},
  publisher={Elsevier}
}

@article{zapata2025enhancing,
  title={Enhancing sparse index-tracking portfolios using deep learning models},
  author={Zapata Quimbayo, Carlos Andres and Arag{\'o}n Urrego, Daniel and Moreno Trujillo, John Freddy and Reyes Nieto, Oscar Eduardo},
  journal={SN Computer Science},
  volume={6},
  number={3},
  pages={199},
  year={2025},
  publisher={Springer}
}

@article{talaei2025robust,
  title={A robust possibilistic programming approach for selecting a clustering-based index tracking portfolio},
  author={Talaei, Mohammad and Torshizi, Ehsan},
  journal={Expert Systems with Applications},
  volume={279},
  pages={127545},
  year={2025},
  publisher={Elsevier}
}

@article{dhingra2026comprehensive,
  title={A comprehensive review and analysis of different modeling approaches for financial index tracking problem},
  author={Dhingra, Vrinda and Sharma, Amita and Goel, Anubha},
  journal={arXiv preprint arXiv:2601.03927},
  year={2026}
}

@article{xu2026network,
  title={Network-based index tracking using asset dependency structures},
  author={Xu, Fengmin and Li, Benchu and Ma, Jieao and Li, Xuepeng},
  journal={International Transactions in Operational Research},
  volume={33},
  number={3},
  pages={1498--1524},
  year={2026},
  publisher={Wiley Online Library}
}

@article{grassetti2025optimizing,
  title={Optimizing index tracking: A Random Matrix Theory approach to portfolio selection},
  author={Grassetti, Francesca},
  journal={Physica A: Statistical Mechanics and its Applications},
  volume={674},
  pages={130747},
  year={2025},
  publisher={Elsevier}
}

@article{zhang2025index,
  title={Index tracking via sparse bayesian regression and collaborative neurodynamic optimization},
  author={Zhang, Fangyu and Wang, Jun},
  journal={IEEE Transactions on Cybernetics},
  volume={55},
  number={3},
  pages={1238--1249},
  year={2025},
  publisher={IEEE}
}

@article{chang2025index,
  title={Index-tracking rigidity and arbitrage opportunities in MSCI index reconstitutions},
  author={Chang, Xin and Luo, Jiang and Peng, Jiaxin and Qian, Shuoge and Tan, Choon Wee},
  journal={Pacific-Basin Finance Journal},
  pages={102900},
  year={2025},
  publisher={Elsevier}
}

@article{de2024assessing,
  title={Assessing the interactions amongst index tracking model formulations and genetic algorithm approaches with different rebalancing strategies},
  author={De Amorim, Thiago Wanderley and Silva, Julio Cezar Soares and de Almeida Filho, Adiel Teixeira},
  journal={Soft Computing},
  volume={28},
  number={6},
  pages={4847--4860},
  year={2024},
  publisher={Springer}
}

@article{silva2023systematic2,
  title={A systematic literature review on solution approaches for the index tracking problem},
  author={Silva, Julio Cezar Soares and de Almeida Filho, Adiel Teixeira},
  journal={IMA Journal of Management Mathematics},
  pages={dpad007},
  year={2023},
  publisher={Oxford University Press}
}

@article{roll1992mean,
  title={A mean/variance analysis of tracking error},
  author={Roll, Richard},
  journal={The Journal of Portfolio Management},
  volume={18},
  number={4},
  pages={13--22},
  year={1992},
  publisher={Institutional Investor Journals Umbrella}
}

@article{jorion2003portfolio,
  title={Portfolio optimization with tracking-error constraints},
  author={Jorion, Philippe},
  journal={Financial Analysts Journal},
  volume={59},
  number={5},
  pages={70--82},
  year={2003},
  publisher={Taylor \& Francis}
}

@book{cornuejols2006optimization,
  title={Optimization methods in finance},
  author={Cornuejols, Gerard and T{\"u}t{\"u}nc{\"u}, Reha},
  volume={5},
  year={2006},
  publisher={Cambridge University Press}
}

@article{jansen2002optimal,
  title={Optimal benchmark tracking with small portfolios},
  author={Jansen, Roel and Van Dijk, Ronald},
  journal={Journal of Portfolio Management},
  volume={28},
  number={2},
  pages={33},
  year={2002},
  publisher={Pageant Media}
}

@article{beasley2003evolutionary,
  title={An evolutionary heuristic for the index tracking problem},
  author={Chang, T-J},
  journal={European Journal of Operational Research},
  volume={148},
  number={3},
  pages={621--643},
  year={2003},
  publisher={Elsevier}
}

@article{derigs2004local,
  title={On a local-search heuristic for a class of tracking error minimization problems in portfolio management},
  author={Derigs, Ulrich and Nickel, Nils-H},
  journal={Annals of Operations Research},
  volume={131},
  pages={45--77},
  year={2004},
  publisher={Springer}
}

@article{gaivoronski2005optimal,
  title={Optimal portfolio selection and dynamic benchmark tracking},
  author={Gaivoronski, Alexei A and Krylov, Sergiy and Van der Wijst, Nico},
  journal={European Journal of operational research},
  volume={163},
  number={1},
  pages={115--131},
  year={2005},
  publisher={Elsevier}
}

@article{alexander2005indexing,
  title={Indexing and statistical arbitrage},
  author={Alexander, Carol and Dimitriu, Anca},
  journal={The Journal of Portfolio Management},
  volume={31},
  number={2},
  pages={50--63},
  year={2005},
  publisher={Institutional Investor Journals Umbrella}
}

@article{oh2005using,
  title={Using genetic algorithm to support portfolio optimization for index fund management},
  author={Oh, Kyong Joo and Kim, Tae Yoon and Min, Sungky},
  journal={Expert Systems with applications},
  volume={28},
  number={2},
  pages={371--379},
  year={2005},
  publisher={Elsevier}
}

@article{maringer2007index,
  title={Index tracking with constrained portfolios},
  author={Maringer, Dietmar and Oyewumi, Olufemi},
  journal={Intelligent Systems in Accounting, Finance \& Management: International Journal},
  volume={15},
  number={1-2},
  pages={57--71},
  year={2007},
  publisher={Wiley Online Library}
}

@article{canakgoz2009mixed,
  title={Mixed-integer programming approaches for index tracking and enhanced indexation},
  author={Canakgoz, Nilgun A and Beasley, John E},
  journal={European Journal of Operational Research},
  volume={196},
  number={1},
  pages={384--399},
  year={2009},
  publisher={Elsevier}
}

@article{ruiz2009hybrid,
  title={A hybrid optimization approach to index tracking},
  author={Ruiz-Torrubiano, Rub{\'e}n and Su{\'a}rez, Alberto},
  journal={Annals of Operations Research},
  volume={166},
  pages={57--71},
  year={2009},
  publisher={Springer}
}

@article{krink2009differential,
  title={Differential evolution and combinatorial search for constrained index-tracking},
  author={Krink, Thiemo and Mittnik, Stefan and Paterlini, Sandra},
  journal={Annals of Operations Research},
  volume={172},
  pages={153--176},
  year={2009},
  publisher={Springer}
}

@article{stoyan2010two,
  title={A two-stage stochastic mixed-integer programming approach to the index tracking problem},
  author={Stoyan, Stephen J and Kwon, Roy H},
  journal={Optimization and Engineering},
  volume={11},
  number={2},
  pages={247--275},
  year={2010},
  publisher={Springer}
}

@article{guastaroba2012kernel,
  title={Kernel search: An application to the index tracking problem},
  author={Guastaroba, Gianfranco and Speranza, Maria Grazia},
  journal={European Journal of Operational Research},
  volume={217},
  number={1},
  pages={54--68},
  year={2012},
  publisher={Elsevier}
}

@article{beasley1990or,
  title={OR-Library: distributing test problems by electronic mail},
  author={Beasley, John E},
  journal={Journal of the operational research society},
  volume={41},
  number={11},
  pages={1069--1072},
  year={1990},
  publisher={Taylor \& Francis}
}

@article{chen2012robust,
  title={Robust portfolio selection for index tracking},
  author={Chen, Chen and Kwon, Roy H},
  journal={Computers \& Operations Research},
  volume={39},
  number={4},
  pages={829--837},
  year={2012},
  publisher={Elsevier}
}

@article{wang2012mixed,
  title={A mixed 0--1 LP for index tracking problem with CVaR risk constraints},
  author={Wang, Meihua and Xu, Chengxian and Xu, Fengmin and Xue, Hongang},
  journal={Annals of Operations Research},
  volume={196},
  pages={591--609},
  year={2012},
  publisher={Springer}
}

@article{ni2013stock,
  title={Stock index tracking by Pareto efficient genetic algorithm},
  author={Ni, He and Wang, Yongqiao},
  journal={Applied Soft Computing},
  volume={13},
  number={12},
  pages={4519--4535},
  year={2013},
  publisher={Elsevier}
}

@article{scozzari2013exact,
  title={Exact and heuristic approaches for the index tracking problem with UCITS constraints},
  author={Scozzari, Andrea and Tardella, Fabio and Paterlini, Sandra and Krink, Thiemo},
  journal={Annals of Operations Research},
  volume={205},
  pages={235--250},
  year={2013},
  publisher={Springer}
}

@article{bruni2015linear,
  title={A linear risk-return model for enhanced indexation in portfolio optimization},
  author={Bruni, Renato and Cesarone, Francesco and Scozzari, Andrea and Tardella, Fabio},
  journal={OR spectrum},
  volume={37},
  number={3},
  pages={735--759},
  year={2015},
  publisher={Springer}
}

@article{guastaroba2016linear,
  title={Linear programming models based on omega ratio for the enhanced index tracking problem},
  author={Guastaroba, Gianfranco and Mansini, Renata and Ogryczak, Wlodzimierz and Speranza, Maria Grazia},
  journal={European Journal of Operational Research},
  volume={251},
  number={3},
  pages={938--956},
  year={2016},
  publisher={Elsevier}
}

@article{filippi2016heuristic,
  title={A heuristic framework for the bi-objective enhanced index tracking problem},
  author={Filippi, Carlo and Guastaroba, Gianfranco and Speranza, Maria Grazia},
  journal={Omega},
  volume={65},
  pages={122--137},
  year={2016},
  publisher={Elsevier}
}

@article{strub2018optimal,
  title={Optimal construction and rebalancing of index-tracking portfolios},
  author={Strub, Oliver and Baumann, Philipp},
  journal={European journal of operational research},
  volume={264},
  number={1},
  pages={370--387},
  year={2018},
  publisher={Elsevier}
}

@article{sant2017index,
  title={Index tracking with controlled number of assets using a hybrid heuristic combining genetic algorithm and non-linear programming},
  author={Sant’Anna, Leonardo Riegel and Filomena, Tiago Pascoal and Guedes, Pablo Cristini and Borenstein, Denis},
  journal={Annals of Operations Research},
  volume={258},
  pages={849--867},
  year={2017},
  publisher={Springer}
}

@article{wu2017constrained,
  title={A constrained cluster-based approach for tracking the S\&P 500 index},
  author={Wu, Dexiang and Kwon, Roy H and Costa, Giorgio},
  journal={International Journal of Production Economics},
  volume={193},
  pages={222--243},
  year={2017},
  publisher={Elsevier}
}

@article{strub2019two,
  title={A two-stage approach to the UCITS-constrained index-tracking problem},
  author={Strub, Oliver and Trautmann, Norbert},
  journal={Computers \& operations research},
  volume={103},
  pages={167--183},
  year={2019},
  publisher={Elsevier}
}

@article{sehgal2019enhanced,
  title={Enhanced indexing using weighted conditional value at risk},
  author={Sehgal, Ruchika and Mehra, Aparna},
  journal={Annals of Operations Research},
  volume={280},
  pages={211--240},
  year={2019},
  publisher={Springer}
}

@article{gnagi2020tracking,
  title={Tracking and outperforming large stock-market indices},
  author={Gn{\"a}gi, Mario and Strub, Oliver},
  journal={Omega},
  volume={90},
  pages={101999},
  year={2020},
  publisher={Elsevier}
}

@article{kaucic2020polynomial,
  title={Polynomial goal programming and particle swarm optimization for enhanced indexation},
  author={Kaucic, Massimiliano and Barbini, Fabrizio and Camerota Verd{\`u}, Federico Julian},
  journal={Soft Computing},
  volume={24},
  number={12},
  pages={8535--8551},
  year={2020},
  publisher={Springer}
}

@article{sant2020solving,
  title={Solving the index tracking problem based on a convex reformulation for cointegration},
  author={Sant'Anna, Leonardo Riegel and de Oliveira, Alan Delgado and Filomena, Tiago Pascoal and Caldeira, Jo{\~a}o Frois},
  journal={Finance Research Letters},
  volume={37},
  pages={101356},
  year={2020},
  publisher={Elsevier}
}

@article{sant2022risk,
  title={Risk measure index tracking model},
  author={Sant’Anna, Leonardo Riegel and Righi, Marcelo Brutti and M{\"u}ller, Fernanda Maria and Guedes, Pablo Cristini},
  journal={International Review of Economics \& Finance},
  volume={80},
  pages={361--383},
  year={2022},
  publisher={Elsevier}
}

@article{torri2023penalized,
  title={Penalized enhanced portfolio replication with asymmetric deviation measures},
  author={Torri, Gabriele and Giacometti, Rosella and Paterlini, Sandra},
  journal={Annals of Operations Research},
  pages={1--51},
  year={2023},
  publisher={Springer}
}

@article{silva2023using,
  title={Using GAN-generated market simulations to guide genetic algorithms in index tracking optimization},
  author={Silva, Julio Cezar Soares and de Almeida Filho, Adiel Teixeira},
  journal={Applied Soft Computing},
  volume={145},
  pages={110587},
  year={2023},
  publisher={Elsevier}
}

@article{vieira2023liquidity,
  title={Liquidity-constrained index tracking optimization models},
  author={Vieira, Eduardo Bered Fernandes and Filomena, Tiago Pascoal and Sant’anna, Leonardo Riegel and Lejeune, Miguel A},
  journal={Annals of Operations Research},
  volume={330},
  number={1},
  pages={73--118},
  year={2023},
  publisher={Springer}
}

@article{anis2023risk,
  title={Risk-allocation-based index tracking},
  author={Anis, Hassan T and Costa, Giorgio and Kwon, Roy H},
  journal={Computers \& Operations Research},
  volume={154},
  pages={106219},
  year={2023},
  publisher={Elsevier}
}

@article{beraldi2022enhanced,
  title={Enhanced indexation via chance constraints},
  author={Beraldi, Patrizia and Bruni, Maria Elena},
  journal={Operational Research},
  pages={1--21},
  year={2022},
  publisher={Springer}
}

@article{silva2024enhanced,
  title={An enhanced grasp approach for the index tracking problem},
  author={Silva, Julio Cezar Soares and Silva, Diogo Ferreira de Lima and de Almeida Filho, Adiel Teixeira},
  journal={International Transactions in Operational Research},
  volume={31},
  number={3},
  pages={1828--1858},
  year={2024},
  publisher={Wiley Online Library}
}

@article{bertsimas2003robust,
  title={Robust discrete optimization and network flows},
  author={Bertsimas, Dimitris and Sim, Melvyn},
  journal={Mathematical programming},
  volume={98},
  number={1-3},
  pages={49--71},
  year={2003},
  publisher={Springer}
}

\pagebreak

\end{document}